\documentclass{iopart}
\usepackage{iopams}
\usepackage{graphicx,xcolor}
\usepackage[utf8]{inputenc}
\usepackage[backend=bibtex, sorting=none]{biblatex}
\usepackage{sidecap}
\usepackage{lipsum}
\usepackage[perpage]{footmisc}

\newcommand{\Pfus}{\ensuremath{P_\mathrm{fus}}}
\newcommand{\Paux}{\ensuremath{P_\mathrm{aux}}}
\newcommand{\Pbrem}{\ensuremath{P_\mathrm{B}}}
\newcommand{\fsa}[1]{\left\langle{#1}\right\rangle}
\newcommand{\eqref}[1]{(\ref{#1})}
\newcommand{\rev}[1]{{\color{black}#1}}

\begin{document}

\title{Physics design point of high-field stellarator reactors}

\author{J~A~Alonso,  I~Calvo,  D~Carralero,  J~L~Velasco, J~M~García-Regaña, I Palermo and D  Rapisarda}
\address{Laboratorio Nacional de Fusi\'on, CIEMAT, Av. Complutense 40, 28040 Madrid}
\ead{arturo.alonso@ciemat.es}

\ioptwocol

\begin{abstract}
The ongoing development of electromagnets based on High Temperature Superconductors has led to the conceptual exploration of high-magnetic-field fusion reactors of the tokamak type, operating at on-axis fields above 10 T. In this work we explore the consequences of the potential future availability of high-field three-dimensional  electromagnets on the physics design point of a stellarator reactor. 
We find that, when an increase in the magnetic field strength $B$ is used to maximally reduce the device linear size $R\sim B^{-4/3}$ (with otherwise fixed magnetic geometry), the physics design point is largely independent of the chosen field strength/device size. A similar degree of optimization is to be imposed on the magnetohydrodynamic, transport and fast ion confinement properties of the magnetic configuration of that family of reactor design points. Additionally, we show that the family shares an invariant operation map of fusion power output as a function of the auxiliary power and relative density variation. \rev{The effects of magnetic field over-engineering and the $R(B)$ scaling of design points with constant neutron wall loading are also inspected. In this study we use geometric parameters characteristic of the \emph{helias} reactor, but most results apply to other stellarator configurations}.
\end{abstract}

\section{Introduction}

In magnetic confinement fusion, the stellarator configuration offers viable solutions for what today appear to be serious problems of the more developed tokamak concept. Since the confining field of a stellarator is created by external coils, without the need to drive large amounts of current within the plasma, continuous operation is granted.  This improves the plant availability factor, which positively impacts the overall tritium breeding ratio \cite{AbdouNF2021}. Furthermore, since the intrinsic plasma current in a stellarator is typically small, sudden plasma terminations are not expected to limit the reliability or the integrity of a stellarator reactor device. Major disruptions, causing large electromagnetic forces in the structure of the device and generating very energetic beams of run-away electrons, continues to be a co	ncern for tokamak reactors. 

The confinement laws of both tokamaks and stellarators display a gyro-Bohm dependence \cite{DinklageFST2007}. However, contrary to tokamaks, $H$-mode operation is not considered in baseline stellarator reactor designs.  The understanding of $H$-mode transition in stellarators is not any more mature than in tokamaks, but observations suggest that a strong pedestal formation is not typical of $H$-modes in stellarators, {with temperature profiles showing no clear edge pedestal}. On the one hand, this leads to only mild improvements in confinement for those regimes ($\sim20\%$) and, on the other, to a more benign edge localised mode activity \cite{ErckmannPRL1993, HirschPPCF2008, EstradaCPP2010}. Operating tokamak reactors in $H$-mode imposes a minimum power across the separatrix which, in combination with the requirement to radiate a large fraction of the power that enters the scrape-off layer, has been shown to possibly limit the design space of a fusion reactor based on the tokamak concept \cite{ReinkeNF2017, SiccinioNF2017}.

The heat exhaust problem, i.e. the need to handle the vast amount of plasma self-heating as it eventually crosses the magnetic separatrix and travels towards the divertor plates, is shared by all approaches to magnetic confinement fusion. The greater diversity of stellarator configurations results in a more diversified scrape-off layer and divertor concepts and geometries. The helical coils of the heliotron device offer a natural long-legged divertor with a well separated volume \cite{OhyabuNF1994}. The helical axis advanced stellarator (\emph{helias}) relies on the so-called island divertor, which features multiple X-points and long connection lengths.  The operation of an island divertor requires to reduce the plasma current by design, and/or actively control it, to position the field resonance in the plasma periphery. Quasi-axisymmetric and helically-symmetric stellarators feature large bootstrap currents and therefore need to resort to `non-resonant' divertors \cite[see e.g.][]{BoozerJPP2015}, that, so far, have not found an experimental realisation\footnote{The helically symmetric experiment (HSX) has been shown to possess a resilient localised pattern of field-line strike points \cite{BaderPoP2017}, but a divertor was not foreseen in its construction.}. 
The island divertor concept was first implemented in the Wendestein 7-AS stellarator \cite{SardeiJNM1997} and further engineered in its successor, Wendelstein 7-X (W7-X) \cite{HermannFST2004}. First W7-X divertor operation has demonstrated a potential to handle power  \cite{NiemannNF2020} and impurities \cite{WintersPhD2019} favourably. Furthermore, long and stable power detachment has been attained \cite{ZahngPRL2019, SchmitzNF2020}. The characterisation and understanding of the island divertor scrape-off layer physics and plasma wall interaction \cite{MayerPS2020} will occupy the fusion community for years to come, but significant progress is already being made \cite{KillerPPCF2020, FengNF2021, BrezinsekNF2021}.

In this and other respects, the stellarator physics basis is under development and  the engineering feasibility of a stellarator power plant needs to be demonstrated. The three-dimensional magnetic configuration poses specific challenges for confining the hot plasma fuel and fusion-generated fast particles. It also complicates engineering aspects, which mainly arise from the three-dimensionality of the electromagnets and the need to have enough of them, close enough to the plasma, to generate an optimised magnetic configuration. Superconducting coil manufacture and support structure, breeding blanket geometry or remote maintenance are all more challenging in stellarator reactors \cite{WarmerFSD2017}. Nevertheless, advances in coil optimization codes are providing new insights and less constraining geometries \cite{LandremanPoP2016, YamaguchiNF2019}.

In the scientific literature, there exist stellarator reactor studies based on several magnetic configurations, including compact quasi-symmetric \cite{NajmabadiFST2008}, force-free heliotron \cite{GotoPFR2011, SagaraNF2017} and helias \cite{GriegerFT1992, BeidlerNF2001, WarmerFST2015}. An early comparison of them was published in \cite{BeidlerToki2001}. \rev{Recently, the fusion reactor systems code PROCESS has been adapted to deal will stellarator configurations \cite{LionNF2021}, which will allow to assess economic and technological feasibility of specific 3D equilibria and coil sets.}

The physics model used in reactor studies is generally rather coarse-grained, such that only the general geometric parameters of the different configurations (e.g. aspect ratio, rotational transform, size and magnetic field) enter the description through scalings and operational limits inferred from experience in one or several devices. Similarly, for devising design points for stellarator reactors, assumptions need to be made on aspects such as the shape of the density and temperature profiles, the concentration of helium ash or the efficiency of the alpha particle heating. This is simply the reflection of an incomplete understanding of the transport processes in magnetic confinement plasmas. Current stellarator research aims at filling those gaps, understanding the relevant equilibrium, stability and transport physics that would allow identifying credible design points for a stellarator reactor. 

The present work builds on these studies, and shares their main methods and assumptions, to assess the impact of a coil technology capable of producing strong confining magnetic fields, stronger than those possible with conventional superconductors like niobium-tin, on the physics design point of a stellarator reactor. The demonstration of the fundamental technological principle, namely, the high temperature superconductors (HTS) of the REBCO type, exist since several decades \cite{FiskSSC1987}, but has only recently received a broader attention from the tokamak  (see e.g. \cite{SorbomFED2015, WhyteJFE2016, MumgaardAPS2017}) \rev{and stellarator \cite{BrombergFST2011, Paz-SoldanJPP2020} fusion communities (see also the recent review \cite{BruzzoneNF2018})}. As a first step towards the demonstration of the high-field path to commercial fusion energy, researchers have set off to demonstrate the technological feasibility of high-field  HTS magnets for its later use in small-scale proof-of-principles experimental devices \cite{CreelyJPP2020}. 

Since both the strength of the magnetic field and device size positively affect the energy confinement time, a trade-off between them is possible. Reference \cite{ZohmPTRSA2019} concludes that the ability to produce stronger fields in tokamak configurations (of the order of 10 T on-axis) allows to reduce the size ($\sim$5 m major radius), and potentially also the cost, of a fusion reactor while maintaining sufficient confinement for ignition. In doing so, an obvious problem arises: a smaller wall surface leads to larger neutron wall loads and a faster degradation of in-vessel components. This requires not only alternative neutron stopping and tritium breeding approaches, like those based on molten-salt liquid blankets, but also \rev{developments in maintenance schemes (e.g. demountable coils)} that reduce the reactor down-time  \cite{WhyteJFE2016}. {The increase in field and current density on the coils yields larger stresses on the structural components, which can pose important engineering challenges.} 
Furthermore, the unmitigated parallel heat flux towards the divertor is also expected to grow as the magnetic field increases (lowering the power flux perpendicular decay length) and the size is reduced (lowering the linear dimension of the strike lines). 

But reducing the linear scale of a stellarator reactor carries additional complications besides those related to higher wall fluxes. The radial extent of the tritium breeding blanket cannot be downscaled, so that the magnetic configuration of smaller devices need to be produced by coils that are proportionally further away from the last closed magnetic surface. Flux surface shaping and tailoring of the magnetic field spectrum is consequently more challenging and leads to strongly shaped coil designs, since the higher-order modes of the $B$-field decay quickly when moving away from the coils \cite{KuNF2010, LandremanPoP2016}. One is led to conclude that the exploitation of HTS for fusion applications must be accompanied by important developments in the design and engineering of a stellarator reactor. These developments could certainly build on the above-mentioned proof-of-principle solutions in the tokamak line. \rev{We note that stellarator reactors based on catalysed D-D fusion and low-temperature superconductors have also been studied \cite{SheffieldFST2016}.} 

In the present work we address the questions: How would very high field HTS-based electromagnets impact the \emph{physics} design point of a deuterium-tritium stellarator reactor? What stellarator physics research directions could prepare us for making use of such a technological development? We make note that it is a premise of this study that the many engineering challenges associated with the fabrication of strongly shaped electromagnets and their operation at very-high fields are surmounted. Although we discuss some engineering aspects of higher-field, smaller devices, the focus of this study lies on the physics characteristics of the reactor design point. 

The rest of the paper is organised as follows. In section \ref{sec:formulas} we present the device geometry, physics assumptions, and formulas that will be used in the one-dimensional study of stellarator reactor design points.  A simplified zero-dimensional analysis is first conducted in section \ref{sec:0D}, where we derive the basic field strength / device size relation that stems from the empirical scaling of energy confinement time. We show how that relation leads to the approximate invariance of several important physical parameters for a family of reactor design points. Importantly, the scaling of the density operation point is introduced in this section (and further elaborated in \ref{sec:density}). In section \ref{sec:1D} we complement the findings of the previous section with 1D profile analysis based on prescribed profile shapes. The 0D invariances translate into several archetypal plasma profiles that are shown and discussed. The effect of magnetic field over-engineering on the physics design point is also discussed in this section. Our main conclusions are summarised in section \ref{sec:conclusions}.    
    
\section{Identification of stellarator reactor design points in the $(B, R)$ plane}\label{sec:formulas}

In this section we present the basic parameters, formulas and hypotheses that are used to identify potential reactor design points in the ($B, R$) plane. Here $B$ is the characteristic  on-axis magnetic field strength and $R$ is the major radius of the device. We will be working with a fixed device aspect ratio and magnetic geometry, so that $R$ characterises the device size and determines all other dimensions such as the minor radius of the plasma column, $a$, or the    confinement volume $V_a = 2\pi^2Ra^2$. 

Given certain values for $B$ and $R$, and other confinement-relevant configuration parameters, we would like to find plasma parameters that fulfil the simplified version of the power balance equation,
\begin{equation}\label{eq:PB}
 P_h = \frac{W}{\tau_E}~,
\end{equation}
where $P_h$ is the net heating power, $\tau_E$ the energy confinement time and $W$ the total internal plasma energy given by the volume integral
\begin{equation}\label{eq:W}
W =  \frac{3}{2}\sum_{s}\int d^3\mathbf{x}\; n_sT_s~,
\end{equation}
where the integration domain is the volume within the last closed magnetic surface.
The index $s$ labels the plasma species with particle density $n_s$ and temperature $T_s$. In the rest of the paper, we will assume that all species have the same temperature and therefore drop the species index and refer to the plasma temperature $T$. Plasma is composed of electrons, with number density $n_e$, deuterium and tritium ions and helium ash from the DT fusion reactions. We assume equal amounts of D and T and a 5\% He concentration, i.e. $n_\mathrm{He}=0.05n_e$, so that $n_\mathrm{D}=n_\mathrm{T}= 0.45n_e$\footnote{We will later show that this concentration of helium implies a ratio of He particle to energy confinement time of about 9 for the family of reactors discussed in section \ref{sec:1D}.}.

The energy confinement time, $\tau_E$ in equation \eqref{eq:PB} is assumed to be given by the ISS04 scaling \cite{YamadaNF2005},
\begin{equation}\label{eq:iss04}
\tau_E = f_c \times 0.134 a^{2.28}R^{0.64}P_h^{-0.61}\overline{n}_e^{0.54}B^{0.84}\iota_{2/3}^{0.41}\; ,
\end{equation}
where $f_c$ is the configuration factor, $\overline{n}_e$ is the line-averaged electron density and $\iota_{2/3}$ is the value of the rotational transform at the $\rho=2/3$ magnetic surface. The normalised radius $\rho$ as well as the radius $r$  will be used to label magnetic surfaces in this work, which relate to each other and to the volume within the flux surface, $V$, as $\rho = r/a = \sqrt{V/V_a}$.  In terms of these variables, a volume integral, like the one in \eqref{eq:W}, is given by
\begin{equation*}
	\int_0^{V_a} dV\; f(V) = 2V_a\int_0^1d\rho\; \rho f(\rho) =  \frac{2V_a}{a^2}\int_0^adr\; r f(r)~,
\end{equation*}
whereas the line-averaged electron density in \eqref{eq:iss04} is given $\bar{n}_e = \int_0^1 n_e(\rho) d\rho$. Note that all physical magnitudes are assumed constant on flux surfaces in this study.

\begin{figure}
	\includegraphics{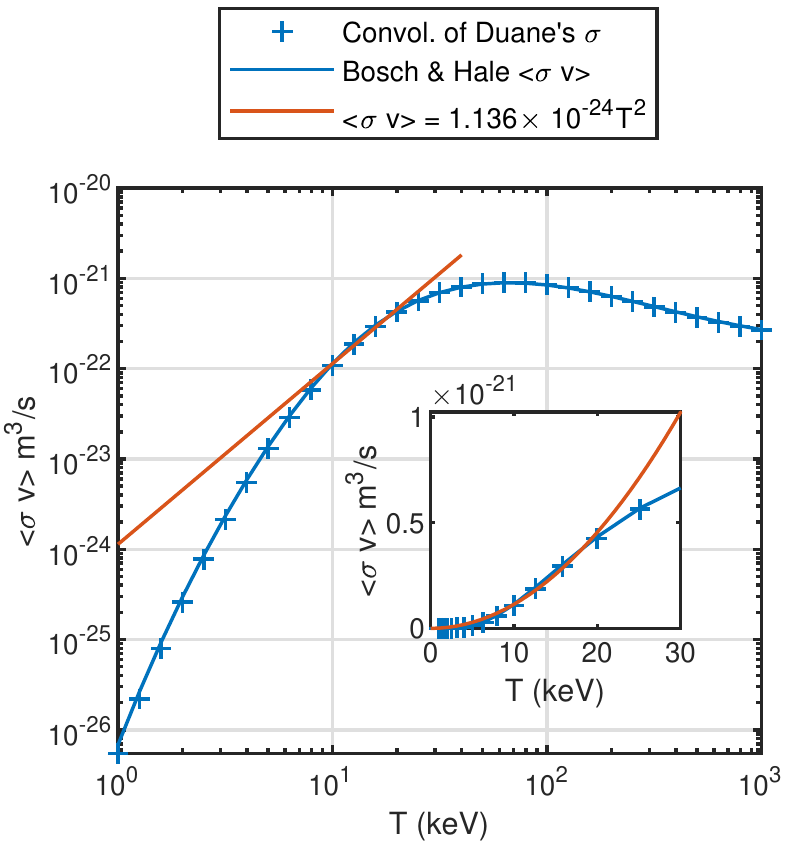}
	\caption{Reactivity of the D-T fusion reaction as a function of the fuel temperature. The points (+) are calculated by convolving  the Maxwellian distributions with the cross section $\sigma$ given by Duane's parametrization \cite{nrl}. The blue curve is the parametric fit to the reactivity given by Bosch and Hale \cite{BoschNF1992}. The red parabola approximates the reaction fairly well below 20 keV (the multiplying coefficient is chosen to match the reactivity at 10 keV).\label{fig:DT}}
\end{figure}

For estimating the net heating power, $P_h$, we consider the plasma self-heating by alpha particles, the power lost by Bremsstrahlung radiation and any auxiliary external heating power. Each of these terms is modelled with a $\rho$-dependent power density, $S_\alpha, S_B$ and $S_\mathrm{aux}$ that are given below:
 \begin{equation}\label{eq:Salpha}
S_\alpha = E_\alpha n_D n_T \fsa{\sigma v}_{DT}(T)~,
\end{equation}
with $E_\alpha = 3.5$ MeV. The form of the DT reactivity, $\fsa{\sigma v}_{DT}$, can be seen in figure \ref{fig:DT}. It displays an approximate quadratic temperature dependence in the range of interest. 
The local Bremsstrahlung radiation density is given by
\begin{equation}\label{eq:SB}
S_B= C_BZ_\mathrm{eff} n_e^2 \sqrt{T}~.
\end{equation}
If temperature and density are expressed in keV and $10^{20}$ m$^{-3}$ respectively, the numerical constant $C_B =5.35\times 10^{-3}$ gives the Bremsstrahlung radiation density in units of MW/m$^3$. The effective ion charge $Z_\mathrm{eff}$ is defined as $Z_\mathrm{eff} = \sum_{s\neq e}Z_s^2n_s/ \sum_{s\neq e}Z_s n_s$ and is equal to 1.1 with our assumed plasma composition. The auxiliary power is modelled by a centred Gaussian 
\begin{equation}\label{eq:Saux}
S_\mathrm{aux} = \frac{P_\mathrm{aux}}{V_a w^2} e^{-\frac{\rho^2}{w^2}}~,
\end{equation}
where $P_\mathrm{aux}$ is the total auxiliary power and $w$ is the width of the power deposition profile. The volume integrals of the power densities \eqref{eq:Salpha} and \eqref{eq:SB} are termed $P_\alpha$ and $P_B$ respectively.

The net heating power is then defined here as $P_h = \max(P(\rho))$, where
\begin{equation}\label{eq:P}
P(\rho)= 2V_a\int_0^{\rho} d\rho\;  \rho(k_\alpha S_\alpha(\rho) + S_\mathrm{aux}(\rho) - S_B(\rho))~.
\end{equation}
The $k_\alpha$ constant is the efficiency of the alpha heating. We take $k_\alpha = 0.9$, thereby allowing for a 10\% loss alpha particle energy due to fast ion transport\footnote{Lower orbit losses are achievable through optimization in time-independent fields \cite[see e.g. the recent study in][]{BaderJPP2019}, but turbulence and/or Alfvén modes can enhance fast particle transport. The specific choice of $k_\alpha$ is not important for the conclusions of this work, as far as it stays sufficiently close to 1.}. 

For the purpose of this study, a reactor design point will be a point in the $(B, R)$ plane for which a device producing a fusion power $P_\mathrm{fus} = 3$ GW with a fusion gain $Q=P_\mathrm{fus}/P_\mathrm{aux} = 40$\footnote{Again, this arbitrary choice does not affect our conclusions. The reactor design point does not depend strongly on the desired $Q$, for large enough $Q$ values (see figure \ref{fig:scan}.j). {Lowering the required auxiliary power would improve the balance of plant but could make active burn control by density necessary.} \rev{Similarly, the lower recirculating power of stellarator power plants could justify targeting lower $P_\mathrm{fus}$ values (see e.g. \cite{MenardNF2011}). We choose to stay with 3 GW to allow a more straightforward comparison with other reactor design points in the stellarator literature.}} can be conceived, on the basis of the presently known scaling of the energy confinement and the density limit. The 3 GW total fusion power is typical of fusion reactor studies and results in an electrical power of about 1 GW, similar to that of present-day fission reactors. A power plant delivering substantially more than 1 GW to the network would have a correspondingly higher capital cost and a more difficult integration in a national power grid. In this work, the total fusion power, $P_\mathrm{fus}$, includes the contribution of the energetic alpha particles and neutrons resulting from the DT reactions, as well as the exothermic neutron-lithium reactions in the breeder \cite{FreidbergPoP2015, RubelJFE2019}. This results in a ratio of fusion to alpha power $P_\mathrm{fus}/P_\alpha = 6.4$.

\begin{table}
\begin{tabular}{p{0.57\columnwidth}cc}
Parameter name & Notation & Value\\
\hline
Aspect ratio & $A$ & 9.0\\
Rotational transform ($\rho =2/3$) & $\iota_{2/3}$ & 0.9\\
Configuration factor & $f_c$ & 1.0\\
Helium concentration & $n_\mathrm{He}/n_e$ & 0.05\\
Fuel concentration & $n_{D,T}/n_e$ & 0.45\\
Effective charge & $Z_\mathrm{eff}$ & 1.1\\
$\alpha$-heating efficiency & $k_\alpha$ & 0.9\\
Fusion-to-alpha power ratio & $P_\mathrm{fus}/P_\alpha$ & 6.4\\
\hline
\end{tabular}
\caption{Reactor design parameters kept fix in this study. Note that the fusion power accounts for alpha and neutron energy, as well as for exothermic breeder reactions \cite{FreidbergPoP2015, RubelJFE2019}. \label{tab:params}}
\end{table}
As mentioned at the beginning of this section, the magnetic geometry will be held fixed in this work. For the formulas presented before, this reduces to fixing the aspect ratio ($A=R/a$), $\iota_{2/3}$ and the configuration factor $f_c$. For the first two magnitudes, we choose values that are representative of the Helias Stellarator Reactor HSR4/18 ($A=9.0$, $\iota_{2/3}=0.9$). The configuration factor, $f_c$, that enters the evaluation of the energy confinement time equation \eqref{eq:iss04} will be set to 1.0 throughout this study. These and other choices are summarised in table \ref{tab:params}. 

In the next sections we will be using these equations (or simplified versions of them) to find potential reactor design points and characterise their physical parameters. We note that, since the alpha heating power depends on the plasma temperature, equations \eqref{eq:PB} and \eqref{eq:W} need to be iterated until they converge to a consistent power balance. 
 
\section{Simplified 0D analysis: $B(R)$ scaling of a reactor design point and its main plasma parameters}\label{sec:0D}

Before presenting the analysis of reactor design points based on the formulas introduced in the previous section, we carry out next a simplified 0D analysis that will guide intuition and help understand those results. We will identify the $B(R)$ dependence of a  reactor design point and the ensuing dependency of the main physics dimensionless parameters on $B$ (or $R$). In order to do that, we will introduce an \emph{ad hoc} scaling of the plasma density that will also be used later in this article.

For the 0D analysis of this section, we rewrite equation \eqref{eq:W} using $\sum_s n_s \approx 2n_e$  and defining $\fsa{n_eT}_V = \frac{1}{V_a}\int_0^{V_a}dV (n_eT)$, to get
\begin{equation}\label{eq:PB0D}
	\fsa{n_eT}_V \approx \frac{P_h\tau_E}{3V_a}~.
\end{equation}
A 0D plasma beta is then defined as
\begin{equation}
\beta_0 = \frac{2\fsa{n_eT}_V}{B^2/2\mu_0}~,
\end{equation}
whereas a characteristic temperature is obtained from the relation
\begin{equation}\label{eq:T0D}
T_{0} = \frac{\fsa{n_eT}_V}{\overline{n}_e}~,
\end{equation}
which is assumed to be close to the volume-averaged temperature. This temperature is used to estimate the alpha and the Bremsstrahlung power. In equation \eqref{eq:PB0D} we include the alpha and auxiliary heating in $P_h$,  but otherwise neglect Bremsstrahlung radiation losses\footnote{In a 0D treatment like the one used in this section, the use of an average temperature like  \eqref{eq:T0D} leads to an overestimation of $P_B/P_\alpha$.}. 

To proceed with the analysis one needs to determine how line-averaged electron density, $\overline{n}_e$, scales with $R$ and $B$. In this work we choose density to scale like
\begin{equation}\label{eq:density}
	\overline{n}_e [\mathrm{m}^{-3}] = 1.04\times 10^{19} (B[\mathrm{T}])^2~,
\end{equation}
where the pre-factor is chosen to match the HSR4/18 density ($2.6\times 10^{20}$m$^{-3}$ at 5T). This choice, which fundamentally determines the conclusions of this article, is consistent with the increase of the cut-off densities of the electron-cyclotron resonance heating an it results in a constant ratio of density to the critical density, $\overline{n}_e/\overline{n}_{ec}$, along the lines constant fusion power and gain in the $(B, R)$ plane. Here $\overline{n}_{ec}$ is the line-averaged critical electron density determined by a Sudo-like radiative limit. Further elaboration of this choice is given in \ref{sec:density}.

\begin{figure*}
  \includegraphics{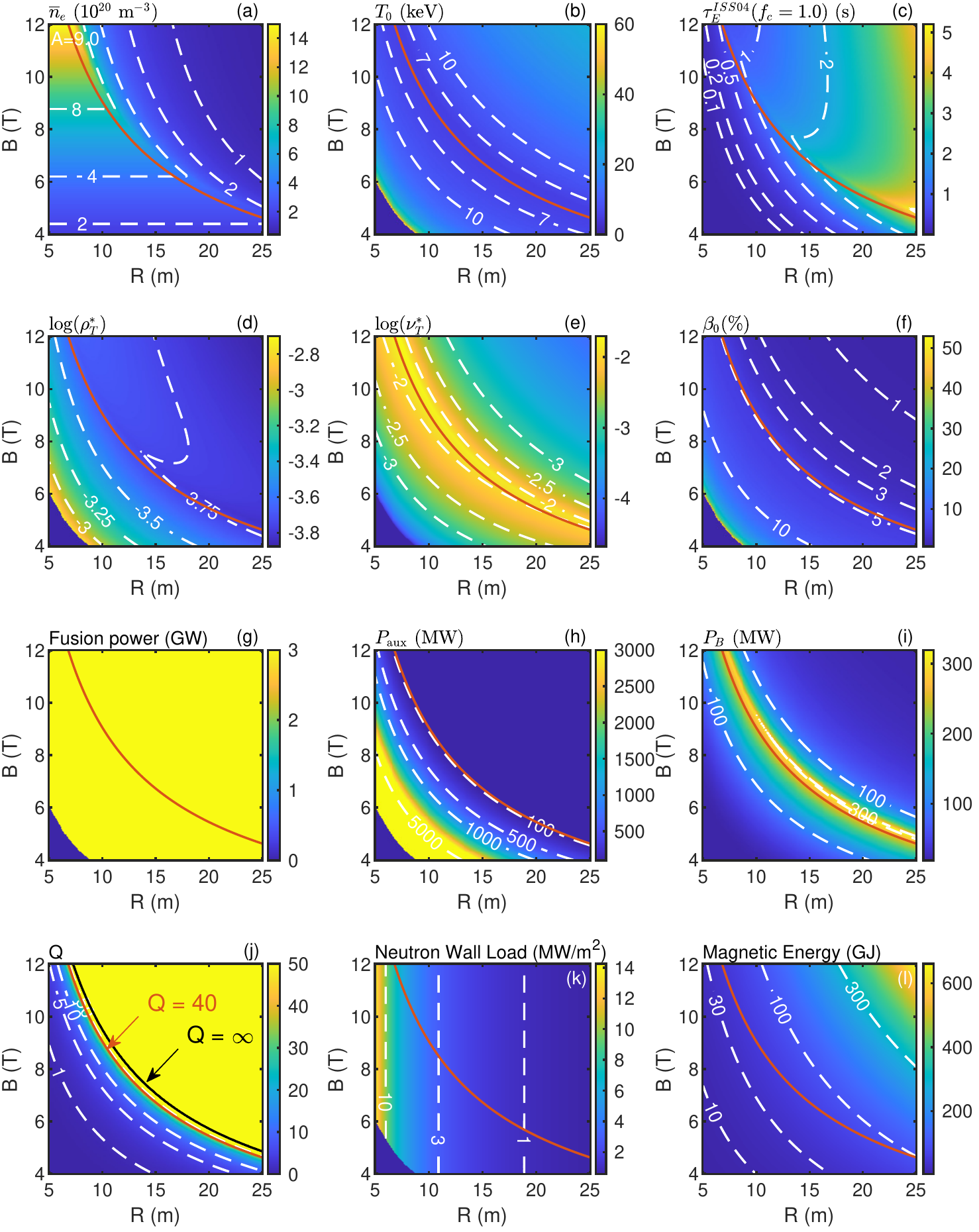}
  \caption{Magnetic field $B$ and device major radius $R$ scan of several physics and engineering parameters. The line-averaged electron density is initially set to the value given by \eqref{eq:density} and the auxiliary power is determined to give a total fusion power of 3GW\rev{, shown in subfigure (g)}. For small, low-field devices this condition cannot be met \rev{(points marked with 0GW in (g))}. For the $Q=\infty$ part of the plane (where $\Paux{} = 0$) the density is lowered to meet the 3GW target. Marked in red is the $Q=40$ line that can be considered for design points of stellarator reactors. We note that the 1D study conducted later in the paper will lower this curve by about 15\%.\label{fig:scan}}
\end{figure*}

The result of the 0D $(B, R)$ scan is shown in figure \ref{fig:scan}.
The plots in this figure are made as follows: with the density initially set to \eqref{eq:density} we proceed, for each $(B, R)$ pair, to find the plasma temperature required to meet the $P_\mathrm{fus} =3$GW target (figure \ref{fig:scan}.g) together with the auxiliary power (figure \ref{fig:scan}.h) that is needed to get to that temperature, according to the $\tau_E$ scaling \eqref{eq:iss04}. For those points for which this process leads to $\Paux{}<0$, the density is lowered until $P_\mathrm{fus} = 3$GW and $\Paux{} = 0$ (this corresponds to the $Q=\infty$ part of the plane). The resulting $(n_e, T)$ pairs are shown in the plots \ref{fig:scan}.a and  \ref{fig:scan}.b. The rest of the physical and engineering parameters (\ref{fig:scan}.c to \ref{fig:scan}.l) are derived from them and the $(B,R)$ values. The definitions of the normalised tritium gyro-radius and collisionality (figures \ref{fig:scan}.d and \ref{fig:scan}.e) are standard and are presented in \ref{sec:rhoandnu} for convenience. 

The consequences of increasing $B$ at a fixed device size while staying with a 3GW fusion target can be seen in figure \ref{fig:scan}: the fusion gain increases rapidly reaching $Q=\infty$. Beyond this point the density operation point needs to be lowered to moderate the increase in the confinement time and reduce the fusion reaction rate. Temperature needs to be increased, which reduces the collisionality. The plasma beta is lowered by the increase in the field strength. While the second of these reductions alleviates the MHD equilibrium and stability requirements on the magnetic configuration, the reduction of density and collisionality makes it more difficult to confine fast and thermal particles in the three dimensional configuration. In consequence we conclude that, unless the reduction of $\beta_0$ design point is a strong requirement, the optimum location of design points for the stellarator reactor in the $B,R$ plane are those just below the $Q=\infty$ line. The red line shown in all plots in figure \ref{fig:scan} are the $Q=40$ $(B, R)$ pairs, which are considered potential reactor design points in this work\footnote{We note that the 1D study conducted later in the paper will shift the $Q=40$ line towards lower fields / smaller device size. In section \ref{sec:Bovereng} we will further discuss the consequences of stepping out of this line.}. 

It is apparent in this figure that points along constant-$Q$ lines (or $Q$-lines) feature a very similar temperature $T_0$ (figure \ref{fig:scan}.b). This is a consequence of the $n\sim B^2$ scaling\footnote{Since all densities are related to each other by constant proportionality factors in this work, we will sometimes use $n$, without subindex, when discussing scalings with either electron or ion densities.}, for the 0D temperature \eqref{eq:T0D} scales as $T_0\sim \tau_E/nV_a$ ($P_h$ held constant). According to \eqref{eq:iss04}, this gives a weak 
\begin{equation}\label{eq:Tconstant}
T_0\sim (RB)^{-0.08}
\end{equation}
dependence. Together with a constant fusion power, $P_\mathrm{fus} \propto n^2V_a\fsa{\sigma v}_{DT}(T)$, this results in the $B(R)$ relation,
\begin{equation}\label{eq:BRrelation}
B \sim R^{-3/4}~, ~(\textrm{constant } P_\mathrm{fus} \textrm{ and } Q)~,
\end{equation}
where the weak dependence \eqref{eq:Tconstant} has been neglected. Note that the aspect ratio is assumed constant throughout this article, so that $V_a\sim R^3$. The $Q=40$ line in the plots in figure \ref{fig:scan} closely follows this $B(R)$ dependence. 

The scaling \eqref{eq:BRrelation} can be understood as the maximal reduction in device size made possible by an increase in the magnetic field strength. That is, given a reactor design point $(R_0, B_0)$, the ability to generate a larger field $B_1$ allows to reduce the relative reactor size  as $R_1/R_0 = (B_0/B_1)^{4/3}$. Further reducing the device size would require to increase the $\overline{n}_e/\overline{n}_{ec}$ ratio to recover the $P_\mathrm{fus} = 3$ GW, $Q=40$ target. It can be shown that the scalings \eqref{eq:Tconstant} and \eqref{eq:BRrelation} stem from the approximate gyro-Bohm dependence of the ISS04 energy confinement time \eqref{eq:iss04}, together with the $n\sim B^2$ scaling used here. In fact, it is straightforward to check that balancing a gyro-Bohm diffusive heat flux, $\Gamma_Q \propto n\chi_{\mathrm{gB}}dT/dr$ (see \ref{sec:chi}), against a heating power divided by the flux-surface area $\sim P_h/R^2$ yields a constant temperature, independent of $B$ and $R$, when the heating power and the normalised temperature scale length are kept constant. The $B(R)$ scaling \eqref{eq:BRrelation} then follows from imposing a constant alpha power with a fixed temperature. 

Several other important physics parameters are kept almost constant along the $Q$-lines in figure \ref{fig:scan}. The Bremsstrahlung power has a dependence similar to that of the alpha power, $\Pbrem{}\sim n^2V_af(T_0)$, where $f(T_0)\sim \sqrt{T_0}$ , and is similarly constant along the $Q$-lines. Plasma beta, $\beta_0 \sim nT_0/B^2$, is approximately constant along the $Q$-lines, whereas triton collisionality ($\nu_T^*\sim nR/T^2\sim R^{-1/2}$) and gyroradius ($\rho_T
^*\sim 1/RB\sim R^{-1/4}$) vary only slowly with $R$ (or $B$). The combination $(\nu_T^*)^{-1}(\rho_T^*)^2$ is approximately constant along the $Q$-lines.  On the contrary, at least two engineering parameters, the neutron wall loading (NWL) and the magnetic energy (figures \ref{fig:scan}.k and  \ref{fig:scan}.l), do show dependencies on the device size. The NWL decreases inversely to the first wall surface area $\sim R^2$ when the neutron power is kept constant, as in this scan. The values shown in the first of these figures have been calculated assuming an average plasma-wall distance required for hosting the diverting magnetic structure equal to 20\% of the plasma minor radius $a$. We defer a longer discussion about the implications of these neutron fluxes and other engineering aspects of high-field devices to section \ref{sec:NWL}. The vacuum magnetic energy is estimated by $\mathcal{E}_B = V_c B^2/2\mu_0$, where the volume enclosed by the coil set of average radii $c$ is approximated by $V_c=2\pi^2Rc^2$. $\mathcal{E}_B$ decreases for smaller, higher-field reactors somewhat slower than $\sim R^3B^2\sim R^{3/2}$ along the $Q=40$ line, since the thickness of the neutron shield and breading blanket is kept equal to 1.3 m in the scan\footnote{\rev{It should be noted that the simplistic estimation of the average coil radius, $c = 1.3 \mathrm{m} + 1.2a$, results in a considerably smaller $c$ and total magnetic energy $\mathcal{E}_B$ than those quoted in \cite{BeidlerNF2001} for $a = 2$ m. Missing a specific MHD equilibrium and coil design, the values and tendencies shown in figure \ref{fig:scan}.l should be considered indicative.}}.

\section{1D analysis of reactor design points with prescribed $(n,  T)$ profiles}\label{sec:1D}
The 0D analysis presented in the previous section allowed to identify dependencies of the main design parameters that are implied by the ISS04 energy confinement time. We found that a quadratic scaling of the density, $n\sim B^2$ leads to a relation $B\sim R^{-3/4}$ of the size and field strength of stellarator reactors with the same fusion energy and gain. This, in turn, leads to the invariance of several physics parameters for a family of reactor design points. However, the precise identification of a design point requires considering profile effects, for most of the fusion reactions are produced in the hottest central part of the plasma column. In this section we will carry out a 1D study of reactor design points and show how the invariances derived in the 0D analysis lead to archetypal profiles for several plasma parameters. Furthermore, we will show that the reactor operation map, $P_\mathrm{fus}(P_\mathrm{aux}, n/n_\mathrm{DP})$, is also shared by the family of stellarator reactors. Here $n_\mathrm{DP}$ is the density at the design point.

The formulas that will be employed for the 1D analysis were presented in section \ref{sec:formulas}, and involve plasma density and temperature profiles. To date, no validated first-principle transport model has been developed that can be used to predict these profiles in a stellarator reactor confidently. Neoclassical transport theory and codes are well established and have been tested, often positively, in 3D magnetic configurations. Neoclassical heat fluxes provide an irreducible transport, found to be generally smaller than the experimental fluxes \cite{DinklageNF2013, BozhenkovNF2020}. Turbulence is thought to cause the additional heat transport, the computation and validation of which is an active subject of current research \cite{BarnesJCP2019, MauerJCP2020, XanthopoulosPRL2020}. An empirical parametrization of turbulent fluxes based on experimental data was used in \cite{WarmerFST2015}. The analysis presented here will be conducted with prescribed shapes of density and temperature profiles. The line-averaged density is given by \eqref{eq:density}, whereas the temperature profile is scaled to provide consistency with the power balance \eqref{eq:PB} and the ISS04 energy confinement time \eqref{eq:iss04}. Details on the form of these profiles are given in \ref{sec:profiles}. We note that the shape of the temperature profile is chosen to depend on the particle and heating power densities with a constant-$\chi$ ansatz. The effect of varying the density profile shape on the reactor design point will be briefly inspected in section \ref{sec:shapes}.
\begin{figure*}
  \includegraphics{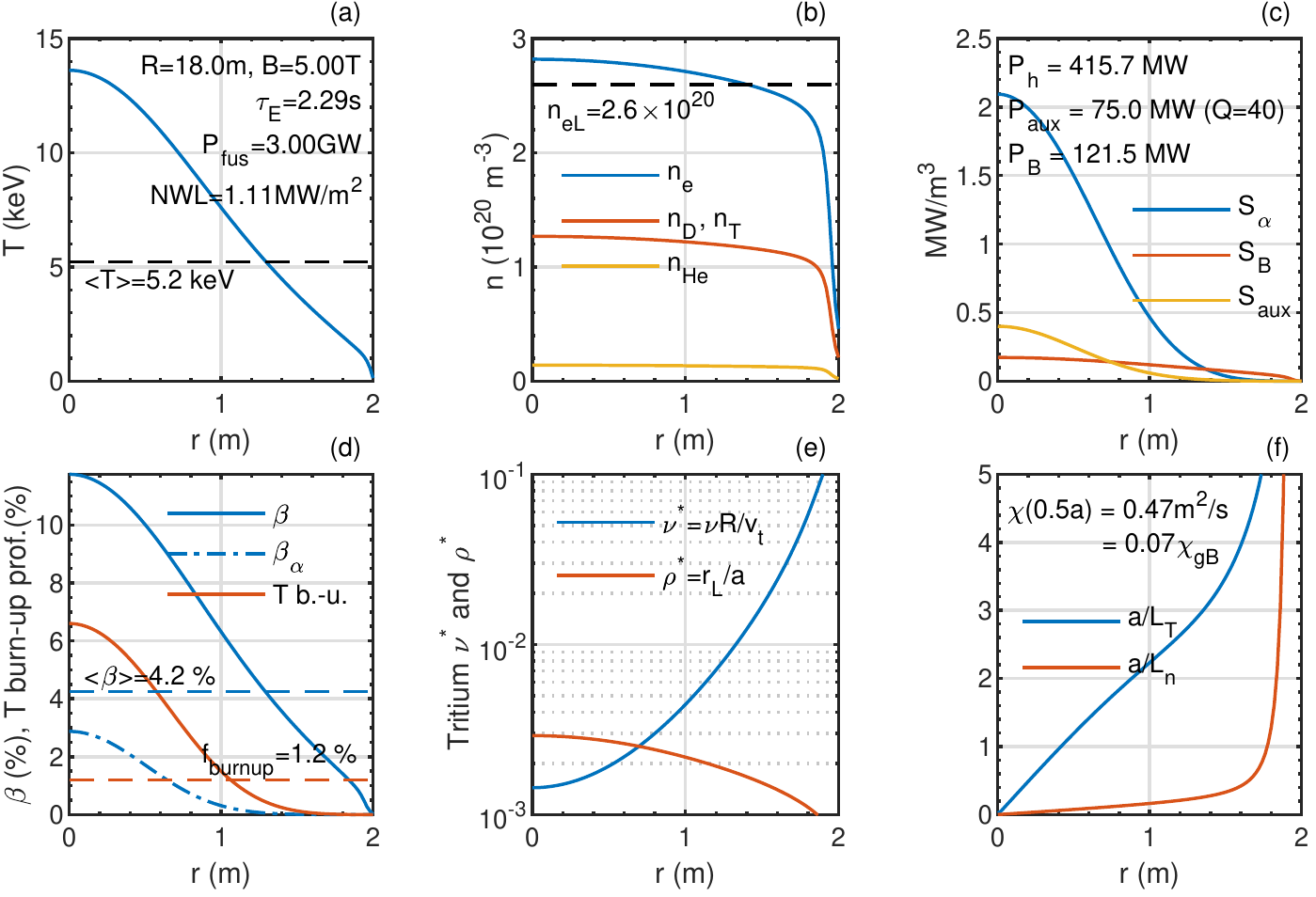}
  \caption{\label{fig:HSR4/18}Plasma profiles for an $R=18$m, $P_\mathrm{fus} = 3$GW, $Q=40$ stellarator reactor. The average values obtained are very close to those of the HSR4/18 (c.f. \cite{BeidlerNF2001}).}
\end{figure*}

An example of the $n$ and $T$ profiles is shown in figure \ref{fig:HSR4/18}.a and b, which approximates the design point of the HSR4/18 device with $R=18$ m, $B=5$ T  and $P_\mathrm{fus} = 3$ GW, $Q=40$. Other plasma parameters, like the volume-averaged plasma beta $\fsa{\beta}$, the energy confinement time $\tau_E$, or the total power radiated through Bremsstrahlung $P_B$, closely resemble design values reported in \cite{BeidlerNF2001}. Note that the scaling of the density \eqref{eq:density} is indeed chosen to match the HRS4/18 line-averaged density at its design field of 5 T.

The resulting alpha heating and Bremsstrahlung power density profiles are shown in figure \ref{fig:HSR4/18}.c, together with the auxiliary heating density of width $w=0.36$ (see equation \eqref{eq:Saux}). Figure \ref{fig:HSR4/18}.d shows the plasma and fast particle beta, $\beta_\alpha$, the tritium burnup profile and total burnup fraction, $f_\mathrm{burnup}$. The classical slowing-down distribution function of alpha particles is used to obtain an estimate for their pressure and beta (see \ref{sec:betaalpha}), which is important to determine the properties of the Alfvén spectrum and the Alfvén-induced energetic particle transport (see e.g. \cite{HeidbrinkPoP2008, } for a review). Recently, a potential stabilising effect of the fast particle pressure on the ion temperature-gradient driven turbulence has been put forward in \cite{DiSienaPRL2020}. 

The tritium burnup fraction is defined as {the number of DT reactions in the plasma per second over the number of injected tritium atoms per second} (see \ref{sec:fburnup}). The inverse of the burnup fraction can be understood as the average number of times that a tritium atom needs to be cycled through the vacuum vessel before it undergoes a DT reaction. Each of the times that a tritium atom is cycled, there is a certain probability that it be lost from the fuel cycle.  The burnup fraction is therefore an important factor to determine the overall tritium breeding ratio. It is, nevertheless, subject to considerable uncertainty. Nishikawa sets a $f_\mathrm{burnup}>0.5$\% requirement to enable tritium self-sufficiency \cite{NishikawaFST2010, NishikawaFST2011}, upon consideration of production of tritium in the blanket system and several losses due to trapping, permeation and decay in the vacuum vessel and fuelling and storage circuits. The same references quote values around 3 - 4\% to ease the requirements on the efficiency of tritium recovery and breeding. A recent review \cite{AbdouNF2021} concludes that tritium self-sufficiency can be achieved with confidence with $f_\mathrm{burnup}>2$\% together with a plant availability factor greater than 50\%. In figure \ref{fig:HSR4/18}.d we estimate the burnup fraction using several, partially compensating approximations; namely, zero recycling, 100\% fuelling efficiency and a equal tritium particle and energy confinement times. This results in an estimated $f_\mathrm{burnup} = 1.2\%$, which falls in the right range\footnote{{Since there is a large uncertainty in this estimate, the quantitative burnup fractions presented in this work should be de-emphasised. More important for the discussion is the fact that, under our assumptions, this fraction does not depend on the reactor size (see below).}}.

%
Finally, fundamental quantities related to the neoclassical and turbulent transport are plotted in figure \ref{fig:HSR4/18}.e (normalised tritium gyro-radius and collisionality profiles) and figure \ref{fig:HSR4/18}.f (temperature and density scale lengths and the thermal diffusivity at mid radius, $\chi_{0.5}$; see \ref{sec:chi}).

\begin{figure}
  \includegraphics{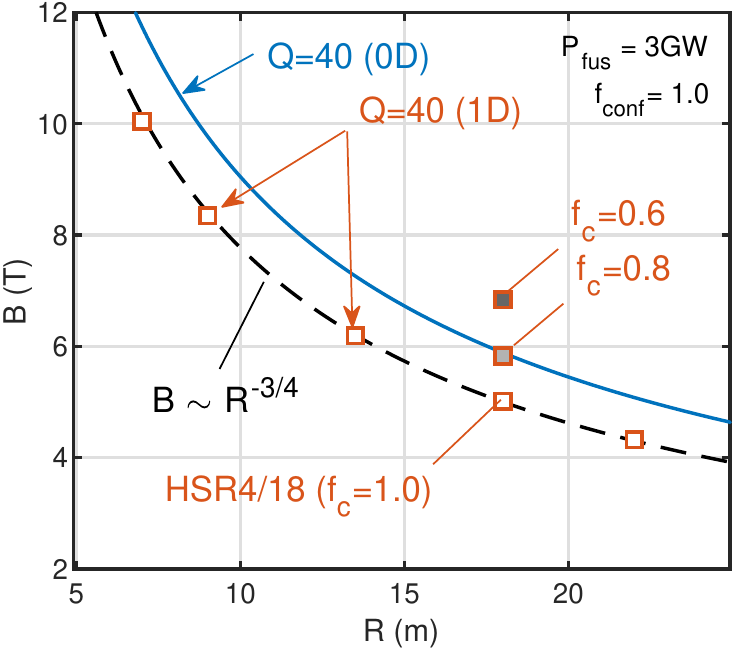}
  \caption{Design point of stellrator reactors in the $B,R$ plane. The 0D analysis conducted in this article overestimates the required field by using an average temperature. The $B\sim R^{-3/4}$ scaling is plotted alongside the $B,R$ pairs obtained from the 1D analysis for various devices sizes $R=$ 7, 9, 13.5, 18 and 22 m (empty squares). Design parameters for the last four of these radii are listed in table \ref{tab:Q40}. These design points share profiles of several parameters shown in figure \ref{fig:invariant} and the operation map shown in figure \ref{fig:OPmap}. The effect of reducing the configuration factor $f_c$ in \eqref{eq:iss04} on the required magnetic field is illustrated for $R=18$m (full squares; note that the line-averaged density for the design points with $f_c = 0.6$ and $0.8$ is still scaled according to $n\sim B^2$, which leads to an increasing ratio of $\overline{n}_e/\overline{n}_{ec}$ for those points).\label{fig:Q40}}   
\end{figure}

The $(B,R)$ position of the design point with profiles shown in figure \ref{fig:HSR4/18}, is plotted in figure \ref{fig:Q40} together with those of smaller, higher-field devices and a larger $R=22$ m device.  
These are calculated similarly, keeping the auxiliary power constant at $P_\mathrm{aux} = 75$ MW  and varying the magnetic field $B$ until the fusion power target, $P_\mathrm{fus} = 3$ GW, is met. As found in the previous section, the resulting design points fall very close to the $B\sim R^{-3/4}$ curve. Table \ref{tab:Q40} summarises the characteristics of those design points. We note that several global magnitudes not listed in that table are nearly constant for the five design points shown in the table (as well as for the other points along the $Q=40$ line in figure \ref{fig:Q40}) as anticipated in section \ref{sec:0D}. This includes the volume averaged plasma $\beta$, the net heating $P_h$ and Bremsstrahlung power $P_B$, the fusion burnup fraction $f_\mathrm{burnup}$ and the ratio of helium particle confinement time $\tau_\mathrm{He}$ to the energy confinement time (see caption in table \ref{tab:Q40}). Furthermore, we will show next that the constancy holds as well for the profiles of certain physics parameters.
\begin{table*}
\begin{tabular}{r r r r r r r r r r}
$R$(m) & $B$(T) & $\overline{n}_e (10^{20}\mathrm{m}^3)$ & $\tau_E$ (s) & $\chi_{0.5}$ (m$^2$/s) & $\nu_T^*(0) 
(10^{-3})$ & W (MJ) & NWL (MW/m$^2$)\\
\hline
7.00 & 10.02 & 10.43 & 0.55 & 0.30 & 2.18 & 223  & 7.32\\
9.00 & 8.32 & 7.20 & 0.80 & 0.34 & 1.95 & 325  & 4.43\\
13.50 & 6.18 & 3.97 & 1.48 & 0.41 & 1.63 & 598  & 1.97\\
18.00 & 5.00 & 2.60 & 2.29 & 0.47 & 1.44 & 922  & 1.11\\
22.00 & 4.31 & 1.93 & 3.10 & 0.52 & 1.32 & 1247 & 0.74\\
\hline
 \end{tabular}
\caption{Parameters of the design points marked in fig.\ref{fig:Q40} with red squares. The parameters $\Pfus{} =3$GW ($P_\alpha = 469$MW) and $Q=40$ ($\Paux{}$ = 75MW) are imposed by our definition of reactor design point. For all points in the table $P_h = 416.5 \pm  1.5$ MW, $P_B = 119.6 \pm  2.9$ MW, $\fsa{T}_V = 5.27 \pm 0.07$ keV, $\fsa{\beta}_V = 4.28 \pm 0.06$\%, $\beta_\alpha(0)/\beta(0) = 0.25 \pm 0.01$, $f_\mathrm{burnup} = 1.21 \pm 0.01$, $\tau_\mathrm{He}/\tau_E = 9.06\pm0.09$. {The helium particle confinement time is estimated using the fusion reactions as the only source.}\label{tab:Q40}}
\end{table*}

At this point we note that, as already stated in \cite{WarmerFST2015}, the synchrotron radiation is not a relevant loss of power for the usual helias reactor design points. This is also the case for the higher field points considered here. In fact the vacuum synchrotron emission scales as $\sim V_aB^2n_eT_e$, which is constant along the $Q=40$ line in figure \ref{fig:Q40}. The plasma opacity factor depends on the density, field and plasma size as $\sim an_e/B$ (see e.g. \cite{Tamor1988}), which slightly increases with the size of the reactor, $\sim R^{1/4}$, along that line. Also note that, in general, the stellarator design points feature higher densities and lower temperatures compared to the tokamak's, which result in substantially lower synchrotron radiation losses for a similar volume and field strength.

%



%

\subsection{Invariant profiles and common optimization requirements}
\begin{figure*}
  \includegraphics{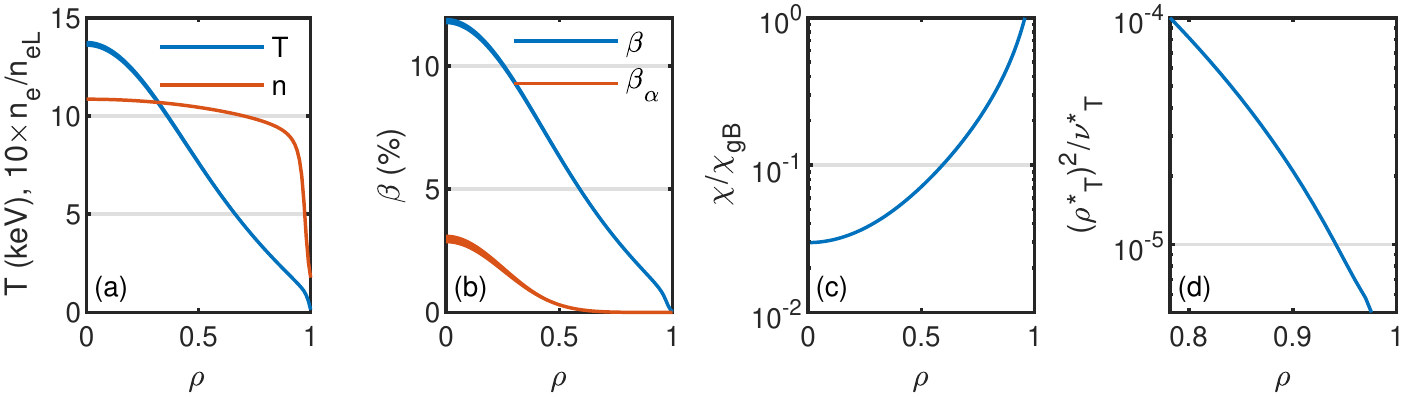}
  \caption{Invariant profiles for the HSR4/18 family (dashed line in figure \ref{fig:Q40}). Line thickness corresponds to the profile variation for the design points listed in table \ref{tab:Q40}.\label{fig:invariant}}
\end{figure*}

There are a number of important plasma profiles that are approximately the same for all the reactor design points with $P_\mathrm{fus}= 3$ GW and $Q=40$ (dashed line in figure \ref{fig:Q40}). This was anticipated in the 0D analysis of section \ref{sec:0D} and is recovered here using a 1D analysis (see figure \ref{fig:invariant}). Shown in this figure are profiles for selected plasma parameters computed for the five design points listed in table \ref{tab:Q40}. Line thicknesses show the maximum differences in the profiles obtained from the 1D power balance described in section \ref{sec:1D}.
These can be considered archetypal profiles for the reactor family that includes the HSR4/18. It should be noted that details of these profiles do depend on specific choices, like the model density and temperature profiles. Their constancy, however, depends solely on the assumption that the same profile shape can be used to represent density for all reactor sizes and on the ISS04 and $n\sim B^2$ scalings. The temperature profile (figure \ref{fig:invariant}.a) is found to be almost constant, and so is the temperature scale-length $a/L_T = T^{-1}dT/d\rho$ ($a/L_n$ is also constant by assumption). Both the thermal and alpha-particle $\beta$ profiles (figure \ref{fig:invariant}.b) and the diffusivity in gyro-Bohm units (\ref{fig:invariant}.c) are nearly invariant. Figure \ref{fig:invariant}.d shows the combination $(\rho^*_T)^2/\nu^*_T$ of normalised gyro-radius $\rho^*_T$ and collisionality $\nu^*_T$. The fact that these profiles are independent of the reactor size (or field) as indicated above, leads to the conclusion that similar levels of optimization of the neoclassical and turbulent transport, MHD equilibrium and stability and fast ion confinement are to be imposed on the magnetic configuration for these design points. We further elaborate on each of these points next.

\subsubsection{MHD optimization.}
That similar equilibrium currents, $\beta$ limits, and MHD stability properties are required is implied by the invariance of the $\beta(\rho)$ profile in figure \ref{fig:invariant}. Modifications to the vacuum magnetic field produced by the plasma current densities $\mathbf{j}$ are given by the equation 
\begin{equation}\label{eq:rotB}
\nabla\times \delta\mathbf{B} = \mu_0\mathbf{j}~.
\end{equation}
The equilibrium plasma currents include the diamagnetic, parallel Pfirsch-Schlüter and parallel bootstrap current, which can be written as
\begin{equation}\label{eq:j}
\mathbf{j} = \frac{dp}{d\rho}\left(\frac{\mathbf{B}\times\nabla\rho}{|\mathbf{B}|^2} + h\mathbf{B}\right) + \frac{\fsa{\mathbf{j}\cdot\mathbf{B}}}{\fsa{|\mathbf{B}|^2}}\mathbf{B}~,
\end{equation}
where $h$ is a function of the space and the magnitude of the magnetic field, defined such that it cancels the total divergence of the term in brackets. This first term, containing the diamagnetic and  Pfirsch-Schlüter currents, scales as $\sim (d\beta/d\rho)B/R$ which, plugged into \eqref{eq:rotB}, leads to $\delta B/B\sim(d\beta/d\rho)$. 
In the absence of external current drive, the second term in \eqref{eq:j} is given by the bootstrap current. While an explicit expression for the bootstrap current, valid for any collisionality regime, does not exist, its asymptotic form in the $1/\nu$ regime \cite{HelanderJPP2017} displays a scaling $\fsa{\mathbf{j}\cdot\mathbf{B}}\sim nT/R$, which, similarly leads to a relative field modification, $\delta B/B\sim\beta$, independent of the device size.  A similar scaling follows from an earlier work by Boozer \cite{BoozerPoFB1990}. Note that the electrostatic potential profile, $\phi(\rho)$, which enters those expressions, is also expected to be nearly invariant, as the neoclassical ambipolarity implies $eT^{-1}(d\phi/d\rho)\sim d\log p/d\rho$, where $e$ is the elementary charge.

\subsubsection{Transport optimization.}\label{sec:transport}
The transport properties of the devices that share the invariant profiles in figure \ref{fig:invariant} should also bear important similarities. 
For all of them, the total thermal diffusivity, $\chi$, must be a similarly small fraction ($\sim 10^{-2}-10^{-1}$) of the gyro-Bohm diffusivity, $\chi_{\mathrm{gB}}$, in the central part of the plasma, which will be shown to be consistent with the expected dependencies in the neoclassical and  turbulent diffusivities. Note that this order of magnitude is imposed by the need to reach a sufficient core temperature to achieve the target fusion power of 3 GW and does not depend strongly on the profiles chosen in this study.

Neoclassical energy transport is deemed an important, possibly dominant, transport mechanism in high-temperature stellarator reactor cores. In 3D fields, it features a specific regime of high diffusivity that is inversely proportional the collisionality. In this so-called $1/\nu$ regime, the heat diffusivity of a plasma species $s$ is $\chi^s_{1/\nu}\propto (a^2v_{ts}/R)(\epsilon_\mathrm{eff}^{3/2}\rho^{*2}_s/\nu_s^*) = (a/R)(\epsilon_\mathrm{eff}^{3/2}/\nu_s^*)\chi_{\mathrm{gB}}^s$, where the proportionality constant is of order unity (see e.g. \cite{SeiwaldJCP2008}). The effective ripple, $\epsilon_\mathrm{eff}$, is a characteristic of the magnetic field structure, the minimisation of which is targeted by stellarator optimization. An order-of-magnitude figure for how small this coefficient needs to be, can be estimated by imposing $\chi \geq \chi_{1/\nu}$ for the tritium ions. Figure \ref{fig:invariant}.c shows that  $\chi/\chi_{\mathrm{gB}}\sim 3\times 10^{-2}$ is necessary in the centre of the family of stellarator reactors. Together with the weakly size-dependent value of the central tritium collisionality in table \ref{tab:Q40} this conditions leads to a central $\epsilon_\mathrm{eff} \lesssim 1\%$. Values around or below this are characteristic of several existing stellarator configurations (see e.g. \cite{BeidlerNF2011} and references therein), including the HSR4/18 \cite{BeidlerNF2001}, new compact quasi-axisymmetric \cite{HennebergNF2019} and several W7-X configurations. In fact, the analysis conducted in \cite{WarmerFST2015} shows that neoclassical transport is compatible with fusion conditions for a scaled version of the high-mirror W7-X configuration. It should also be noted that the radial electric field moderates ion neoclassical losses at the very low central collisionalities, where other regimes like the $\sqrt{\nu}$ become increasingly important. This makes the $1/\nu$ estimates a worst-case scenario for the neoclassical transport channel.

The approximate invariance of scale lengths $a/L_T$, $a/L_n$ and $eT^{-1}(d\phi/d\rho)$, an assumed constant magnetic geometry and the only weak size-dependence of collisionality ($\sim R^{-1/2}$, see the end of section \ref{sec:0D}) and $\rho^*_T\sim R^{-1/4}$ leads to the conclusion that $\chi_\mathrm{neo}/\chi_{\mathrm{gB}}$ would be similar for all reactor design points. Nevertheless, it is well known that microturbulence enhances energy transport above neoclassical levels. In neoclassically optimized stellarators like W7-X, the turbulent component can even dominate the radial energy fluxes over the entire plasma volume \cite{BozhenkovNF2020}. Given a magnetic configuration, $\chi_\mathrm{tb}/\chi_{\rm \mathrm{gB}}$ depends on $a/L_T$, $a/L_n$,and the collisionality. In the family of stellarator reactors discussed in this paper, $a/L_T$, $a/L_n$ do not vary, whereas the collisionality varies only slightly. One is then similarly led to conclude that, $\chi_{\rm tb}/\chi_{\mathrm{gB}}$ would not vary significantly within this family\footnote{\rev{The neoclassical and turbulent diffusivities  $\chi_\mathrm{neo}$ and $\chi_{\rm tb}$ are defined as the coefficients that relate the size of the
corresponding energy flux with the typical scale length of the temperature profile. Namely, $Q_\mathrm{neo}\sim \chi_\mathrm{neo} {nT} L_T^{-1}$ and $Q_\mathrm{tb}\sim \chi_\mathrm{tb} {nT} L_T^{-1}$.}}. 

To conclude the discussion on transport, we note that the fact that $\chi_\mathrm{neo}/\chi_{\mathrm{gB}}$ and $\chi_\mathrm{tb}/\chi_{\mathrm{gB}}$ do not vary much within the reactor family is consistent with the approximate invariance of the required $\chi/\chi_{\mathrm{gB}}$ (figure \ref{fig:invariant}.c). This does not come as a surprise, for the ISS04 energy confinement time displays an approximate gyro-Bohm dependence that is also characteristic of both neoclassical and turbulent transport mechanisms\footnote{This is not to say that deviations in particular parametric dependencies of the neoclassical and turbulence diffusivities with respect to the gyro-Bohm diffusivity are excluded. The neoclassical 1/$\nu$ limit discussed before is an example of this. Collisional stabilization of turbulent trapped electron modes has been shown to possibly lead to an isotope mass dependence inverse to that of gyro-Bohm \cite{NakataPRL2017}. \rev{What is important from the previous discussion is that the other parameters that define the neoclassical and turbulent transport regimes, and might introduce deviations with respect to the gyro-Bohm diffusivity, are themselves nearly equal across the reactor family.}. }.

\subsubsection{Fast ion optimization.}
Collisionless fast ion losses in stellarator proceed in two time-scales. Trapped particles cause prompt losses, which have a very short characteristic time given by $v_M/a$, where $v_M$ is the characteristic size of the radial magnetic drift. Since this  is much shorter than the collisional slowing-down time, the magnetic configuration needs to be designed to reduce prompt losses for any of the reactor sizes considered in figure \ref{fig:Q40}. On a longer time-scale, fast ion losses due to stochastic diffusion set in, which can be moderated by lowering the collisional slowing-down time. A diffusion coefficient for the stochastic losses was derived in  \cite{BeidlerPoP2001} using an analytical representation of the magnetic field spectrum. Aside from size-independent geometric factor, the stochastic diffusion scales as $D_\alpha\sim R^2\Omega_\alpha(\rho^*_\alpha)^4$, where $\Omega_\alpha$ is the alpha particle cyclotron frequency. In terms of our size ($R$) and field strength ($B$) variables, one gets $D_\alpha\sim R^{-2}B^{-3}$, which reflects a reduction in the diffusivity of alpha particles for larger devices and stronger fields. To compare the importance of stochastic losses for the different design points, we use an estimate of the diffusive radial excursion in a slowing-down time , $\fsa{\Delta \rho} \sim a^{-1}\sqrt{D_\alpha \tau_s}$. For constant electron temperature, the slowing-down time $\tau_s$ is inversely proportional to the density (see equation \eqref{eq:slowing-down}). Using the density and $B(R)$ scaling (equations \eqref{eq:density} and \eqref{eq:BRrelation}) leads to $\fsa{\Delta \rho}\sim R^{-1/8}$. By this measure, stochastic diffusion of alpha particles is expected to be of very similar importance for the various reactor design points considered in figure \ref{fig:Q40} and table \ref{tab:Q40}.

Alpha particles can excite Alfvén modes which, in turn, can enhance their radial transport. The characteristics of this interaction and enhanced transport would be similar for the different reactor design points in the following sense: first, the population of alpha particles displays an invariant $\beta_\alpha$ profile shown in figure \ref{fig:invariant} \rev{(see also \ref{sec:betaalpha})}. Second, the Alfvén speed $v_A=B/\sqrt{\mu_0\sum_sm_sn_s}$, where $m_s$ is the mass of the species $s$, is constant according to the density scaling \eqref{eq:density}.

\subsection{Invariant operational map}

\begin{figure}
	 \includegraphics{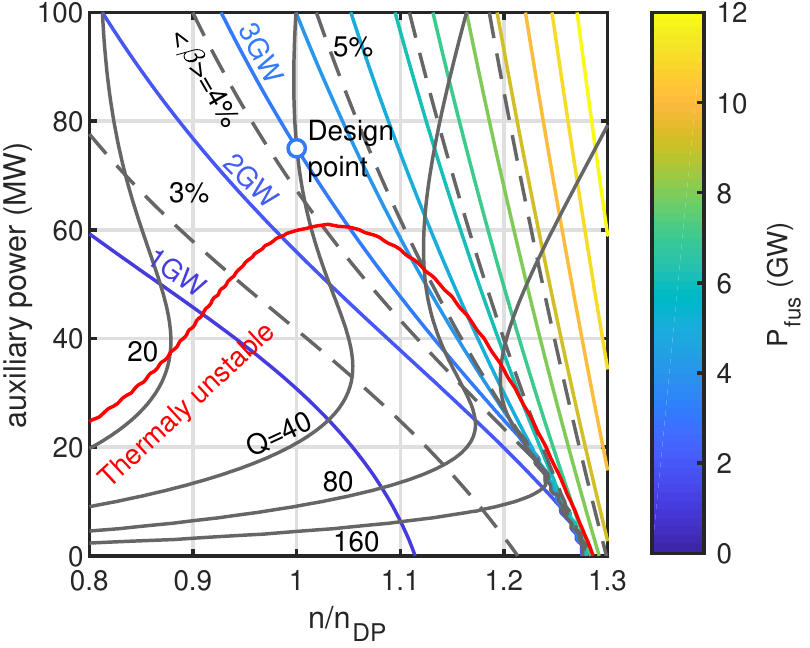}
	 \caption{Operation map ($P_\mathrm{aux}, n/n_{DP}$) for the HSR4/18 reactor family (dashed line in figure \ref{fig:Q40}). Thermal stability refers to the sign of $P_h-W/\tau_E$ caused by a 1\% constant variation of the equilibrium temperature profile. The steep variation of fusion power in the low $P_\mathrm{aux}$, high $n/n_{DP}$ region is due to the appearance of a higher temperature solution of the $P_h = W/\tau_E$ power balance. Note that the fuel mix is kept constant in this scan.\label{fig:OPmap}}
\end{figure}

The family of reactors in figure \ref{fig:Q40} and table \ref{tab:Q40} share another important characteristic; namely, the operation map (i.e. the fusion power dependence on the auxiliary power $P_\mathrm{aux}$ and relative density variation with respect to the design point $n/n_{DP}$), shown in figure \ref{fig:OPmap}.
That this is indeed the case can be shown from the dependences $P_\mathrm{fus} \sim V_a n^2T^2$ and $nT\sim P_h\tau_E/V_a$, which leads to $P_\mathrm{fus}\sim R^{0.005}(n/n_{DP})^{1.08}P_h^{0.78}$. Since the alpha and Bremsstrahlung power within $P_h$ are approximately constant along the $Q$ lines, the operation map $P_\mathrm{fus}(P_\mathrm{aux}, n/n_{DP})$ (figure \ref{fig:OPmap}) is also approximately independent of the reactor size along the constant $Q$ curve (figure \ref{fig:Q40}).  

In figure \ref{fig:OPmap} the thermal stability of the operation points is probed by varying the temperature profile by 1\% and looking at the sign of $P_h - W/\tau_E$, which is, by construction, equal to zero at each operation point for the equilibrium temperature and density profiles. If a positive (negative) increment in the temperature profile leads to a faster increase (decrease) of the heating power, $P_h$, compared to the transported power, $W/\tau_E$, then the operation point is labelled thermally unstable. It is important to note that the shape of the temperature perturbation can affect the sign of the resulting $P_h - W/\tau_E$. Points below the red curve in figure \ref{fig:OPmap} are at least unstable to a temperature perturbation like the one referred above. While operating a stellarator reactor in thermally unstable conditions might be possible, it would presumably require an active control of the burning point for a stable power output. It should also be noted that the way thermal stability is calculated assumes that the energy confinement scales like equation \eqref{eq:iss04} also ``locally'', but deviations from it (including confinement transitions) are observed in present individual devices.

\subsection{Effect of the profile shapes}\label{sec:shapes}
The shape of the density profile has been so far fixed to that shown in figure \ref{fig:invariant}, with a mild peaking given by $k_2=0.5$ in equation \eqref{eq:nemodel}. The shape of the temperature profile \eqref{eq:Tmodel} also depends on this choice. In a reactor, the density profile will be determined by the fuelling method and the particle transport characteristics. Peaked density profiles are thought to be beneficial for confinement in Wendelstein-7X \cite{BozhenkovNF2020}. In the absence of a core particle source,  neoclassical thermodiffusion is however expected to lead to core particle depletion \cite{BeidlerPPCF2018}. Hollow density profiles have not been reported in W7-X so far, but are common in the LHD device.

\begin{SCfigure*}
  \includegraphics{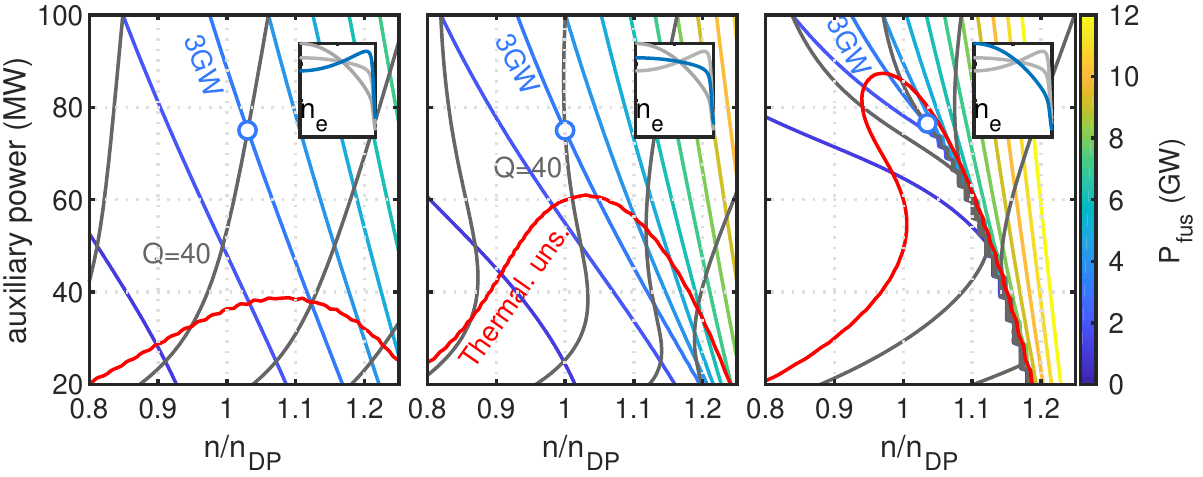}
  \caption{Operation maps for density peaking factors $k_2 = 0.5, -1.0$ and $5.0$ (profiles are shown as insets in each figure) for the HSR4/18 reactor family. Colour curves are contours of constant $P_\mathrm{fus}$ with 1GW increments. Gray curves are contours of constant $Q$ with $\times 2$ increments. Circles mark the intersection of the  $P_\mathrm{fus} = 3$ GW and $Q=$40 lines.\label{fig:peakings}}
\end{SCfigure*}

For inspecting the effect of the profile shape on the operation point of a reactor we reproduce the operation map of figure \ref{fig:OPmap} for a hollow ($k_2 = -1$) and a more peaked  ($k_2=5$) density profiles. This is shown in figure \ref{fig:peakings}, which shows that relatively small adjustments of the auxiliary power and density allow to recover the $P_\mathrm{fus}=3$ GW, $Q=40$ operation point. However, the operation landscape is substantially changed. The sensitivity of $P_\mathrm{fus}$ is greatly increased for the peaked density profile, such that the operation point lies in a thermally unstable region (in the sense described previously) and sits close to the jump to a higher temperature solution of the power balance equation. It needs to be noted that the operation map is calculated assuming that the D-T fuel mix can be kept constant. In practice, the accumulation of He ash would likely lead to the moderation of the reaction rate for the higher fusion powers.

\subsection{Discussion on the neutron wall loads and breeding technology}\label{sec:NWL}
As discussed earlier around table \ref{tab:Q40}, smaller reactor devices inevitably suffer more intense neutron bombarding per unit area on the first wall and breeding blanket. To qualify the neutral wall load (NWL) numbers shown in that table we consider that, as a rule of thumb, a maximum NWL of 1.97 MW/m$^2$ (corresponding to a 13.5 m device) would translate into a damage rate of 19.7 dpa/fpy (displacements per atom in a full-power year) in steel and around $4\times10^{18}$ n/m$^2$/fpy neutron flux at the first wall. This could be still acceptable under the damage of the first wall, shielding of the other structures (Vacuum Vessel and Coils) and heat recovery points of view, as preliminary assessed in \cite{PalermoNF2021}, although some improvement on shielding and minor modification on maintenance scheme would be necessary.
Nevertheless, such aspects would be not easily manageable with higher NWL.  For example, 4.43 MW/m$^2$ ($R=9$ m in table \ref{tab:Q40}) would lead to damage around 40 dpa/fpy and neutron fluxes of almost  $8\times10^{18}$ n/m$^2$/fpy at the first wall. This would complicate  the breeding blanket replacement, since the current Eurofer first wall material suggested to be used in stellarators power plants, as extrapolated from DEMO, is qualified up to 20 dpa for the first DEMO phase (1.57  fpy). Besides, the requirements to the shield components that would need to be very exigent in order to get viable values at the different coil structures On the other hand, the NWL of the largest devices considered in table \ref{tab:Q40}, 0.74 MW/m$^2$ would probably lead to low power deposited inside the coolants and accordingly low thermal efficiency. The fourth case with a NWL of 1.11 MW/m$^2$, which indeed corresponds to the HSR4/18 design point, could be the best engineering solution to explore since it seems to offer the best compromise between damage/shielding performances, maintenance schemes and heat recovery/thermal efficiency. 

Apart from the considerations on the NWL, a substantial increase of the magnetic field can influence other technological aspects related with the in-vessel components. Firstly, important electromagnetic forces can be developed on the structures which form the breeding blanket \cite{MaioneFED2019}. It has been demonstrated that the mechanical behaviour of the blanket segments can be compromised. The breeding material can also be affected by the magnetic field. Usually, Li compounds are required to breed the tritium in order to maintain the reactor self-sufficiency. One of the most extended breeding materials is the PbLi eutectic alloy, which indeed is an excellent electric conductor. This means that, when moving inside the blankets, magnetohydrodynamic effects can appear \cite{UrgorriNF2018}. The main consequence is the increase on the pressure drop of the liquid metal, which has a critical impact on the plant electric efficiency. Moreover, some MHD effects can cause important reverse flows that can compromise the route of the effluent impacting the tritium permeation through the structures or the formation of He clusters. Another important point is corrosion of structural materials due to the interaction with the liquid metal. It has been demonstrated that the presence of an intense magnetic field enhances the corrosion phenomena on EUROFER samples \cite{CarmonaJNM2015}.

The conclusion of this discussion is that current breeding blanket technology and maintenance schemes can severely limit the viability of small-size, high-field reactors. The full exploitation of high temperature superconductors necessitates parallel technological developments on these fronts, that allow one to cope with the increased neutron fluxes. Continuous molten-salt breeders and demountable coil joints, for easier and faster maintenance, have been proposed in the tokamak high-field path to commercial fusion \cite{WhyteJFE2016}. Some of these ideas have also been considered in stellarator reactor studies \cite{SagaraNF2017}.

\subsection{Magnetic field over-engineering}\label{sec:Bovereng}

So far we have centred our analysis on design points on the $Q=40$ line in figure \ref{fig:Q40}. In the 0D analysis of section \ref{sec:0D} we argued that increasing the field over those design points, while keeping the $P_\mathrm{fus} = 3$ GW target, implied to operate at lower densities and higher temperatures. In a 3D device, operating at lower collisionality and longer alpha particle slowing-down times are disadvantages, as they impose higher standards on the neoclassical and fast particle confinement properties of the magnetic configuration. Since the energy confinement time in its usual form \eqref{eq:iss04} has a positive dependence on the line average density, lowering it is, in a sense, a waste of the confinement potential of the magnetic configurations. However, over-engineering the magnetic field strength in the sense just described brings in two positive consequences of large potential impact: it lowers the plasma $\beta$ and increases the tritium burnup fraction. In reduced-size tokamak reactor studies, high field operation is exploited to move away from operational boundaries \cite{WhyteJFE2016}. The increase of the magnetic field and the decrease in the operation density both act to lower the critical density fraction $\overline{n}_e/\overline{n}_{ec}$.  Although it is not in the scope of this article to assess the viability of heat exhaust solutions, we note in passing that those same changes could complicate power handling in the scrape-off layer and divertor regions. 

In this subsection we take a look at the consequences of over-engineering the magnetic field with the 1D prescribed-profile analysis utilised in this section. \rev{We inspect the effect of a 25\% increase in $B$ that for the case of a $R=13.5$ m device (see table \ref{tab:Q40}). The results are shown in figure \ref{fig:Bovereng}. 
\begin{figure}
\includegraphics{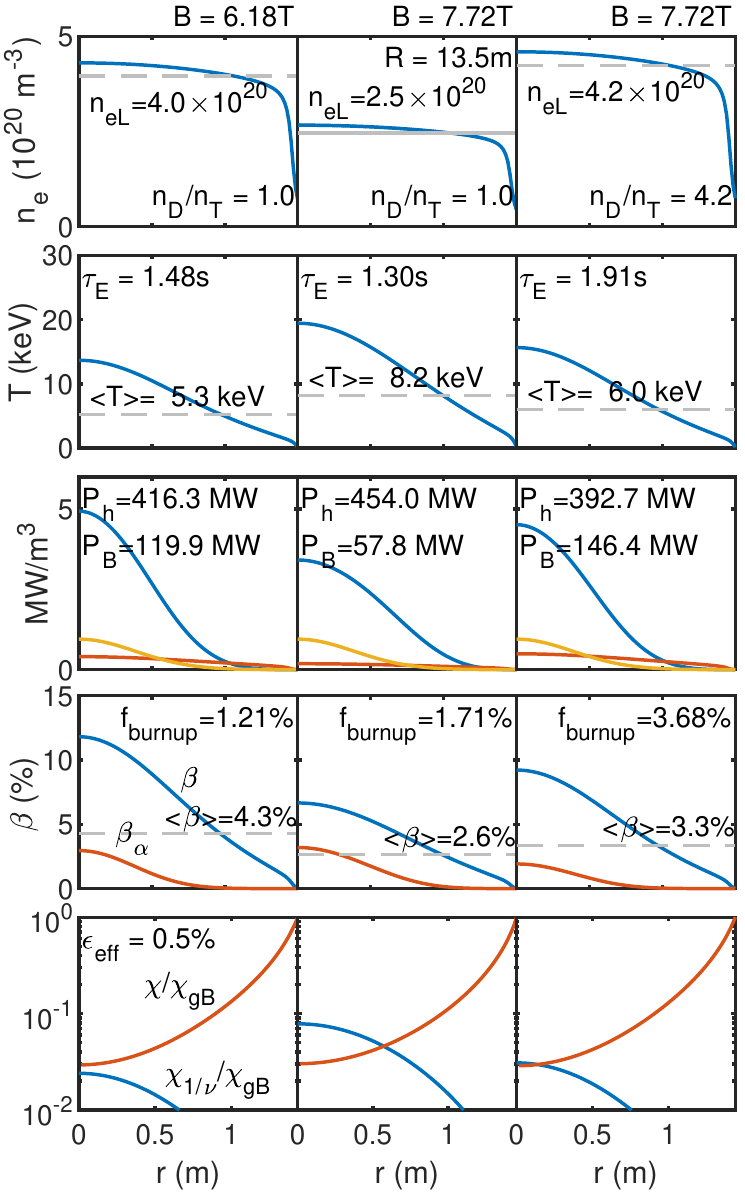}
\caption{Magnetic field over-engineering with respect to the $R=13.5$m design point shown in \ref{fig:Q40}. The design point profiles at $B=6.18$T are shown in the left column. \rev{Middle and right column profiles correspond to a 25\% increase of the magnetic field. In all cases $P_\mathrm{fus}=3$ GW and $P_\mathrm{aux}=75$ MW. In the middle column (with a $n_D/n_T$ density ratio of 1), the density needs to be lowered to moderate the fusion power. In the right column, the density is slightly increased, as allowed by the increase in the critical density (see equation \eqref{eq:nc}), and the $n_D/n_T$ ratio is adjusted instead to recover 3 GW fusion power. The labelling of the power densities in the middle row is the same as in figure \ref{fig:HSR4/18}.c.}\label{fig:Bovereng}}
\end{figure}

The profiles for the starting $R=13.5$ m reactor design point are shown in the left column. Those in the centre and right column correspond to the 25\% over-engineered field case. In the centre profiles, the ratio of deuterium to tritium densities, $n_D/n_T$, is kept at 1. In this situation, the increase in the fusion power due to the increase of $B$ can only be moderated by reducing the plasma density.}  Temperature needs to change in roughly inverse proportion to keep up with the 3 GW fusion power. The higher temperatures broaden the alpha particle generation and heating profile and increase the burnup fraction. The thermal plasma $\beta$ decreases as $B^{-2}$. However, the alpha particle $\beta_\alpha$ slightly increases and broadens as a result of the longer slowing-down time and the broader profile of alpha particle production. These two effects would demand better fast particle confinement properties (in particular for the longer time-scale stochastic diffusion losses) in a more extended radial range for this over-engineered $B$-field case \rev{with $n_D/n_T = 1$}. It should also be noted that for helias-type configurations, a \emph{sufficient} plasma beta is required to reduce fast ion prompt losses, but the necessary minimum beta is subject to some adjustment at the stage of the optimization of the magnetic configuration. The last row of figure  \ref{fig:Bovereng} shows the comparison of the normalised thermal diffusivity $\chi$ (see \ref{sec:chi}) with the proxy for the neoclassical $1/\nu$ diffusivity. The effective ripple is set to $\epsilon_\mathrm{eff} = 0.5\%$. These plots illustrate that further reductions of $\epsilon_\mathrm{eff}$ could be necessary to make the core profiles compatible with the neoclassical heat transport levels in the over-engineered case \rev{with $n_D/n_T = 1$}.

\rev{Instead of lowering the plasma density,  in the right column of figure \ref{fig:Bovereng} the ratio of deuterium to tritium has been increased to moderate the fusion power. This is shown to result in a moderation of the negative consecuences refered to above, that stemmed from the necessary decrease of density and increase of temperature. While the reduction of the plasma $\beta$ is not as pronounced, the $\beta_\alpha$ is also reduced and the collisionality increases only slightly (bottom plot). The reduction of the tritium concentration increases the burnup faction three-fold with respect to the normal field case. We conclude that magnetic field over-engineering could be taken advantage of by reducing the tritium concentration, thereby allowing to reduce the plasma and alpha particle normalised pressures and increase the burnup fraction, without strongly reducing  plasma collisionality.}

\begin{figure}
\includegraphics{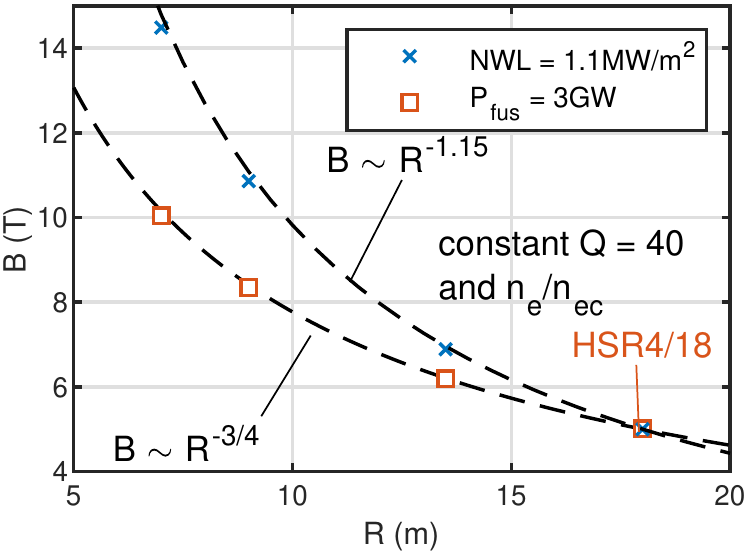}
\caption{\rev{Comparison of reactor design points of power (`$\square$') and NWL (`$\times$') similar to those of the HSR4/18. Note that Q and $\overline{n}_e/\overline{n}_{ec}$ are kept constant for all points. The $B\sim R^{-1.15}$scaling is obtained under the assumption of a nearly invariant temperature profile (see also table \ref{tab:constantNWL}).\label{fig:constantNWL}}}
\end{figure}

\rev{\subsection{Design points of stellarator reactors with constant neutron wall loading}
As mentioned before, the neutron wall loading is an important technological parameter. The breeding blanket and heat-to-electricity transformation technology determine an optimal range for the NWL. Therefore, one could be interested to search for reactor desing points of constant NWL rather than constant fusion power. Reducing the size of the reactor design point while keeping a constant NWL still gives favorable reduction of the ratio of GW per m$^3$ of reactor material, but the larger electro-mechanical stresses of smaller, higher-field devices could make the support structure bigger and more expensive than that implied by a simple $R^3$ scaling.

Referring to the operation map in figure \ref{fig:OPmap}, valid for all reactor design points in table \ref{tab:Q40}, one sees that reducing the auxiliary heating power and adapting the density, one can go down the $Q=40$ line to the fusion power compatible with a given reduced NWL. However, in doing so, one would require the reactor to operate at larger ratios of density over the critical density $\overline{n}_e/\overline{n}_{ec}$, since $\overline{n}_{ec}\sim P_h^{0.57}B^{0.34}R^{-1.25}$. In consequence, if $\overline{n}_e/\overline{n}_{ec}$ is to be kept constant while \emph{reducing} the fusion power at fixed device size, the magnetic field needs to be \emph{increased}. This is a consequence of the strong power dependence of the critical density characteristic of stellarators \cite{GreenwaldPPCF2002}. The resulting $(B, R)$ design points and some of their physics characteristics are shown in \ref{fig:constantNWL} an table \ref{tab:constantNWL} respectively. Besides the critical density fraction, the fusion gain is kept constant at $Q=40$ in those points. As shown in table \ref{tab:constantNWL}, smaller, higher-field devices with constant NWL (i.e. $P_\mathrm{fus} \sim R^{-2}$) display a rapidly decreasing plasma beta, whereas the central collisionality decreases more slowly. The plasma temperature varies only slightly in the explored range.
}
\begin{table*}
\begin{tabular}{r r r r r r r r r}
$R$(m) & $B$(T) & $\overline{n}_e (10^{20}\mathrm{m}^3)$ & $\langle T\rangle$ (keV) & $P_\mathrm{fus}$ (GW) & $\langle\beta \rangle$ (\%) & $\beta(0)/\beta_\alpha(0)$ &  $\nu_T^*(0) (10^{-3})$ & $(\chi/\chi_\mathrm{gB})_{0.5}$\\
\hline
7.00 & 14.58 & 3.89 & 5.54 & 0.45 & 0.79 & 3.68 & 0.77 & 0.05\\
9.00 & 10.86 & 3.58 & 5.34 & 0.75 & 1.27 & 3.91 & 0.96 & 0.06\\
13.50 & 6.89 & 2.98 & 5.25 & 1.69 & 2.58 & 4.04 & 1.23 & 0.07\\
18.00 & 5.00 & 2.60 & 5.21 & 3.00 & 4.24 & 4.10 & 1.44 & 0.07\\
\hline
 \end{tabular}
\caption{\rev{
Reactor design points of constant NWL = 1.11 MW/m$^2$, $\overline{n}_e/\overline{n}_{ec}$ and Q = 40 shown in figure \ref{fig:constantNWL}.\label{tab:constantNWL}}}
\end{table*}

\rev{It should be emphasised that the present understanding of the Sudo-type density limit in stellarators does not allow to determine an absolute limit as is the case for tokamaks' Greenwald limit \cite{GreenwaldPPCF2002}. In this study we have used the published working point of the HSR4/18 stellarator to define the reference $\overline{n}_e/\overline{n}_{ec}$, but the possibility of operating at substantially higher densities cannot be ruled out. If this were to be the case, the magnetic field of the design points could be lowered with respect to those listed in table \ref{tab:constantNWL}. According to the present understanding (see e.g. \cite{FuchertNF2020} and references therein), the density limit in stellarators is connected to the properties and concentration of the edge radiator impurity. Gaining a more predictive capability for the determination of the critical density in reactor conditions is therefore linked with the validation of models of divertor and SOL/edge impurity transport and with the constraints on the concentration of impurities imposed by power exhaust.}

\section{Summary and conclusions}\label{sec:conclusions}

In this article we have shown that the design points of stellarators of different scale and field, but otherwise similar fusion power and gain, share many similarities under the assumption that the plasma density design point can be scaled as $n\sim B^2$. Archetypal profiles for temperature, plasma $\beta$, gradient scale lengths or thermal diffusivity in gyro-Bohm units, among others, characterise a family of varying-field stellarator reactors with maximally-reduced size, $R\sim B^{-4/3}$ (figure \ref{fig:invariant}). The suitability requirements on the magnetic configuration (e.g. good flux surfaces and MHD stability at high-$\beta$, reduced neoclassical transport in the low-collisionality regime or sufficient confinement of alpha particles)  are therefore largely independent of the device size/field strength \rev{across the family of reactor design points}. In this sense, the qualification of optimization criteria in devices like the Wendelstein 7-X is still relevant in a future scenario in which electromagnets based on high temperature superconductors can be applied in stellarator reactors.  Furthermore, we have shown that the operation landscape of fusion power and gain as function of the auxiliary power and relative density is also shared among the family of reactors (figure \ref{fig:OPmap}).  In consequence, if the technological development of high-field-capable electromagnets progresses to make ten-Tesla-class stellarators accessible, a demonstration device of much reduced size ($<$10 m) is conceivable, that would qualify larger-scale devices in several meaningful ways. 

High field and reduced size make conditions particularly harsh for the breeding structure, which calls for alternative approaches for breeder and wall maintenance \cite{WhyteJFE2016}. \rev{An additional complication for devising reduced size stellarator reactor stems from the need to increase the distance from the magnets to the plasma edge (relative to the plasma radius)} to allow for the irreducible space of the breeding blanket and neutron shield. \rev{This results in an increase in the complexity of the coils that would be required to generate an optimised magnetic configuration \cite{LandremanPoP2016}.} It is in this respect that higher-field operation would call for the search of optimised configurations with a larger relative distance between the current and control surfaces. The ability to relax some of the optimization targets, based on possibly over-simplified physics models, would make this line appear more promising. 

If the tritium breeding technology or limitations in remote maintenance were to set the minimum size of a stellarator reactor, one could wish to increase the field strength while maintaining a certain device size. Magnetic field over-engineering is accompanied by reduced thermal $\beta$ and higher tritium burnup fraction, but conditions on thermal and alpha particle confinement become more stringent. \rev{However, these drawbacks can be largely mitigated with the adjustment of the D/T density ratio (figure \ref{fig:Bovereng}).} 

\rev{As an alternative to the constant-$P_\mathrm{fus}$ scaling of reactor design points adopted throughout this article, the scaling at constant neutron wall loading (NWL) and gain has also been inspected in this article. The stellarator-specific critical density dependence on heating  power results in a stronger field scaling ($R\sim B^{-1.15}$) when the reactor linear size is reduced at constant NWL and critical density fraction (figure \ref{fig:constantNWL}). The resulting reactor design points feature lower plasma $\beta$ and slightly lower collisionality compared to same size, higher $P_\mathrm{fus}$ design points.}

The conclusions of this work depend on a number of assumptions that have been presented but cannot be justified on solid physics grounds. The close invariance of plasma profiles and operation map (figures \ref{fig:invariant} and \ref{fig:OPmap}) depends on the use of prescribed shapes for the density and temperature profiles, that are assumed to be the same for all reactors. Other assumptions such as the fuel-mix composition and the alpha heating efficiency add quantitative uncertainty to the design points that have been considered. Our analysis critically relies on the ISS04 scaling of the energy confinement time. The assessment of any limitation to this scaling in high-beta, low-collisionality regimes should be regarded a stellarator research priority. Finally, it should not be left unnoticed that the viability of power exhaust has not been considered in any respect in our study. The development of stellarator scrape-off layer and divertor physics models that allow to incorporate exhaust conditions to reactor studies also appears to be a necessary step in the development of a credible stellarator reactor concept. {Significant departures with respect to the tokamak results \cite{ReinkeNF2017, SiccinioNF2017} could arise from the absence of a well defined threshold power across the separatrix and from differences in the  scaling of the scrape-off layer width \cite{EichNF2013, NiemannNF2020}. \rev{This development bears also the importance of allowing to quantify the Sudo-type density limit that is compatible with exhaust conditions in stellarator reactor studies.}}

\appendix
 
\section{Definitions and conventions} \label{sec:definitions}

\subsection{Model plasma profiles}\label{sec:profiles}

Missing a complete model for particle and energy sources and transport, we fix the shape of the density and temperature profiles used for the 1D analysis. The density profiles is chosen to be represented by 
 \begin{equation}\label{eq:nemodel}
 \hat{n}_\mathrm{model}(\rho) = \frac{\pi}{2} -\mathrm{atan}\left(k_1(\rho^2 - \rho_0^2)\right) - k_2(\rho^2-1)~,
\end{equation}
with $\rho_0^2 =0.95$ (location of the edge density gradient), $k_1=30$ (steepness of the gradient) and $k_2=0.5$ (core flatness). 

Rather than fixing the form of the temperature profile, we choose to make it dependent on the profile of the net heating power density.  We first define the profile function,
\begin{equation}
F(r) = - \int_0^r\frac{1}{r'\sum_sn_s(r')}\left(\int_0^{r'} r''S(r'')dr''\right)dr'~, 
\end{equation}
with $S(r) =k_\alpha S_\alpha+ S_\mathrm{aux} - S_B$. The model temperature profile is expressed in terms of $F(r)$,
\begin{equation}\label{eq:Tmodel}
\hat{T}_\mathrm{model}(r) = 1 + k - \frac{F(r)}{F(a)}~,
\end{equation}
where $k=0.01$ relates to the core to edge temperature ratio, $T(a)/T(0) = k/(1+k)$. The profile given by \eqref{eq:Tmodel} can be seen to result in a radially-constant heat diffusivity.

\subsection{Heat diffusivity}\label{sec:chi}
The total thermal diffusivity is defined as the weighted average of the individual species diffusivities  $\chi = \sum_s \chi_s n_s/\sum_s n_s$. The total heat flux $\Gamma_Q$ is written as $\Gamma_Q = -\frac{3}{2}\chi\sum_s n_s\frac{dT}{dr}$. The steady-state heat balance equation then relates this flux to the heat density $S = k_\alpha S_\alpha + S_\mathrm{aux} - S_B$, 
\begin{equation}
\frac{1}{r}\frac{d}{dr} r\Gamma_Q = S~.
\end{equation}
The diffusivity is calculated as
\begin{equation}
\chi(r) = -\frac{2}{3}\left(r\frac{dT}{dr}\sum_s n_s\right)^{-1}\int_0^r r' S(r') dr'~.
\end{equation}
In some of the plots this is normalised by the local gyro-Bohm diffusivity, defined by $\chi_{\mathrm{gB}} = av_{tT}(\rho_T^*)^2$, where the subindex $T$ refers here to tritium as a plasma species.

\subsection{Normaliised gyro-radius and plasma collisionality}\label{sec:rhoandnu}
The normalsied gyroradius of a plasma species $s$ is denfined in this article as
\begin{equation}
\rho^*_s = \frac{v_{ts}}{\Omega_s a}~,
\end{equation}
where $v_{ts}$ is the thermal velocity $v_{ts}=\sqrt{2T_s/m_s}$ and $\Omega_s$ is the gyrofrequency $\Omega_s = Z_seB/m_s$.

As to the collision frequency of each species, $\nu_s$, we adopt the definition given in \cite[page 3]{BeidlerNF2011}. Collisionality is then given by $\nu_s^* =  \nu_s R/v_{ts}$. For clarity, we only discuss triton collisionality in the main text, which accounts for self-collisions and collisions with deuterons. For equal temperatures, triton collisionality is slightly smaller than deuteron collisionality and about a factor 3 smaller than electron collisionality.

\subsection{$\alpha$-particle pressure}\label{sec:betaalpha}
To estimate the equilibrium alpha particle pressure and beta we use the classical slowing-down distribution function (see e.g. \cite{Helander2002})
\begin{equation}
	f_\alpha(v) = \frac{s_\alpha\tau_s}{4\pi(v^3+v_c^3)}H(v_\alpha-v)~,
\end{equation}
where $s_\alpha = n_Dn_T\fsa{\sigma v}_{DT}(T)$ is the production rate of alphas per unit volume with a birth velocity $v_\alpha$, and $H$ is the Heaviside function. The slowing-down time $\tau_s$ is defined as
\begin{equation}\label{eq:slowing-down}
	\tau_s^{-1} = \frac{4}{3\sqrt{\pi}}\frac{Z_\alpha^2 e^4 n_e \ln\Lambda}{4\pi \epsilon_0^2m_em_\alpha v_{te}^3}~,
\end{equation}
whereas the critical velocity $v_c$ is given by
\begin{equation}\label{eq:vc}
	v_c = \left(\frac{n_iZ_i^2}{n_e}\frac{3\sqrt{\pi}m_e}{m_i}\right)^{1/3}v_{te}~.
\end{equation}
In this equation we consider the deuterium and tritium a single species with 2.5 amu.

The alpha particle pressure is then given by 
\begin{equation}\label{eq:palpha}
p_\alpha = \int d^3\mathbf{v}\;\frac{m_\alpha v^2}{3} f_\alpha(v) = \frac{m_\alpha  v_c^2}{3}s_\alpha\tau_s I\left(\frac{v_\alpha}{v_c}\right)~,
\end{equation}
and
\begin{equation}
I(x) = \int_0^{x}\frac{t^4}{t^3+1} dt~.
\end{equation}
The alpha particle beta, $\beta_\alpha = p_\alpha/(B^2/2\mu_0)$ is a weak function of the reactor size along the $Q=40$ line in figure \ref{fig:Q40}. In fact, assuming a constant temperature (equation \eqref{eq:Tconstant}), the critical velocity \eqref{eq:vc} is constant and \eqref{eq:palpha} scales like $p_\alpha \sim s_\alpha \tau_s\sim n$. With the scaling \eqref{eq:density}, this results in $\beta_\alpha\sim n/B^2 =$ constant.

\subsection{Tritium burnup fraction}\label{sec:fburnup}

The so-called tritium burnup fraction is a global magnitude defined as the ratio between the rate of fusion reactions within the confined volume and the fuelling rate of fresh tritium \cite[see e.g.][]{JacksonFST2013}, namely
\begin{equation}\label{eq:fburnup}
	f_\mathrm{burnup} = \eta_\mathrm{eff}\left( \frac{1}{1 + \frac{N_T}{\tau_T^*K_\alpha}} \right)
\end{equation}
where $\eta_\mathrm{eff}$ is the fuelling efficiency (which depends, in particular, on the fuelling method) and $N_T = \int_VdV\; n_T$ and $K_\alpha = \int_VdV \; n_Tn_D\langle\sigma v\rangle_{DT}$. The \emph{effective} tritium confinement time, $\tau_T^*$, is defined in terms of the tritium confinement time $\tau_T$ and the recycling coefficient\footnote{Note that this coefficient is defined by the fraction of tritons that return to cross the last-closed magnetic surface after having left it. The reactor divertor and scrape-off layer conditions are expected to result in lower recycling values compared to present-day devices, where it can approach unity. In alternative plasma exhaust concepts based on flowing liquid metals, this coefficient would be close to zero.} $R$, as $\tau_T^* = \tau_T/(1-R)$. Having a large $f_\mathrm{burnup}$ reduces the required tritium inventory and losses in a power plant. The figure of merit shown in the figures corresponds to \eqref{eq:fburnup} with $\eta_\mathrm{eff} = 1$, $R = 0$, $\tau_T = \tau_E$ and $N_T = N_D = 0.45N_e$. In figure \ref{fig:HSR4/18} we plot a related quantity labelled tritium burnup profile, that is defined as $\kappa_T = n_D\fsa{\sigma v}_{DT}\tau_E$.

With this definition and assumptions, the tritium burnup fraction is almost constant along the reactor line in figure \ref{fig:Q40} ($P_\mathrm{fus} = 3$ GW, $Q=40$). Since the fusion power is kept constant and is proportional to the alpha power, the $K_\alpha = P_\alpha/E_\alpha$ is constant. The ratio $\tau_E / (N_D + N_T)\sim\tau_E/ (nV_a)$ then scales like plasma temperature in equation \eqref{eq:Tconstant}.

\section{Defining the operation point for density}\label{sec:density}

The choice of density operation point for a reactor bears considerable importance. The fusion power scales approximately as $P_\mathrm{fus}\sim n^2T^2V_a$, so that, for a fixed fusion power, the required temperature scales inversely with density. While the required $\beta$ is roughly independent of the density operation point, the characteristic collisionality ($\nu*\sim n/T^2$) increases quickly with density. The energy confinement time has a positive dependence on density, whereas the alpha particle slowing-down time decreases with it. Furthermore, the divertor operation might require high density operation together with extrinsic radiators in the divertor and {burnup} region for power exhaust. One concludes that the density design point should be made as high as allowed by operation limits and heating/fuelling access. 
\begin{figure}
  \includegraphics{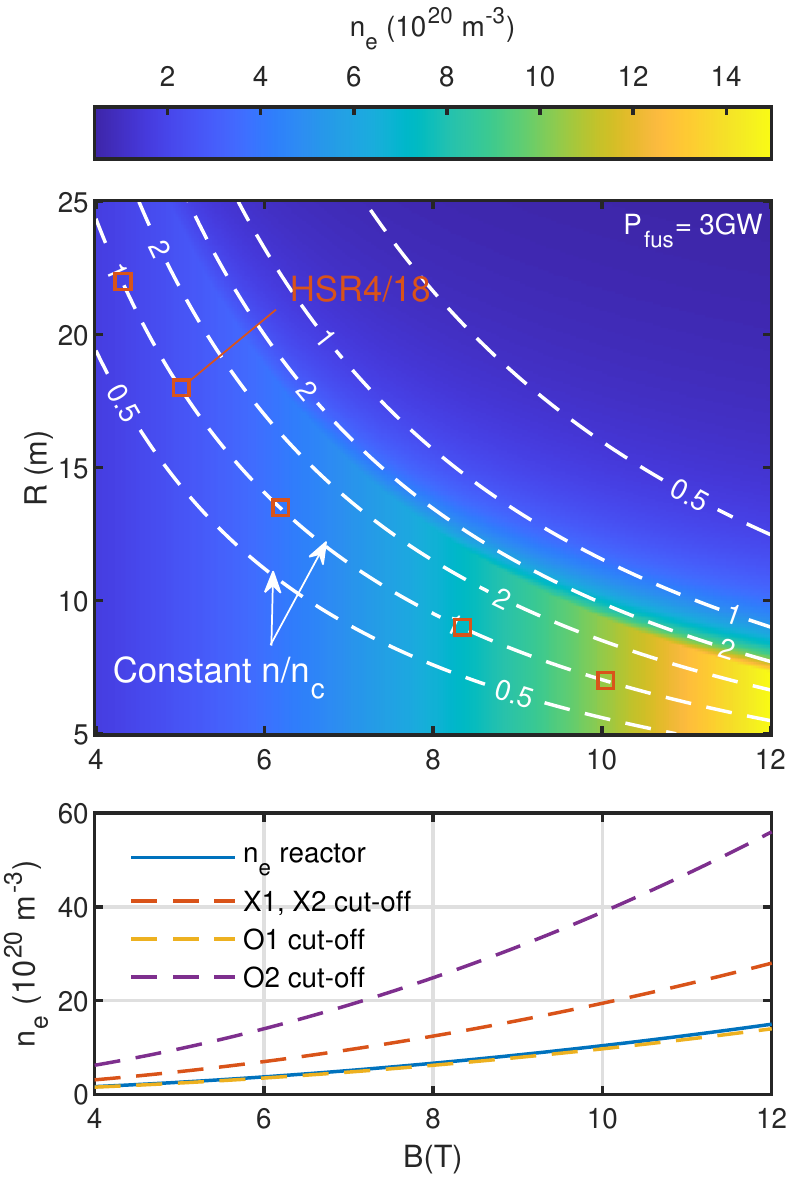}
  \caption{Top: Line-averaged electron density as a function of device size and magnetic field strength as used in the 0D parameter scan in search of reactor design points of section \ref{sec:0D} (see figure \ref{fig:scan}). The dashed lines are contours of the ratio of this density to the density limit in the form \eqref{eq:nc} of reference \cite{FuchertNF2020}. The ratio is normalised to the value for the design point of HSR4/18. This illustrates that the $n\sim B^2$ scaling results in reactor design points (red squares) of constant critical density fraction. Bottom: comparison of the density scaling \eqref{eq:density} with the cutoffs of the electron cyclotron heating in X, and O-modes and first and second harmonic (SX and FX in the legend refer to the slow and fast X-mode waves respectively	). The first-harmonic resonant frequency at 1 T is 28 GHz so $f^N_\mathrm{ECH}[\mathrm{GHz}] = 28\times N\times B[\mathrm{T}]$, where $N=1,2$ respectively for first and second harmonic heating.\label{fig:density}}
\end{figure}

In this study, we have chosen to scale density as $B^2$, which is consistent with the increase of the ECRH cut-off density (see the lower plot in figure \ref{fig:density}). We note that the chosen DP for density is not compatible with having a an O1 ECRH heating scheme. 
Second harmonic fast-X mode or first harmonic slow-X mode (with high-field-side launching) would be required. The gyrotron frequencies that are needed for high field operation have not been achieved to date, so that a technological development would be necessary on this front for operating high field stellarators. In this article we adopt the view that the heating method, whether ECRH or others, should not be the primary density-limiting factor and that such technological developments can be carried out.

The top plot of figure \ref{fig:density} shows that the chosen density scaling is compatible with keeping a constant ratio of density DP to the radiative density limit. The original Sudo density limit \cite{SudoNF1990} for the LHD stellarator is $n_c\sim \sqrt{P_hB/V_a}$, similar to the one found in W7-AS \cite{GiannonePPCF2000}. More recently, the W7-X density limit has been shown to be in agreement with a similar scaling by Fuchert et al. \cite{FuchertNF2020}, namely
\begin{equation}\label{eq:nc}
n_c \propto P_h^{0.57}B^{0.34}R^{-1.25}\;,
\end{equation}
where a constant aspect ratio has been assumed ($a = R/A$). This limit is understood to be due to the critical temperature behaviour of edge impurity radiation \cite{ItohJPSJ1988}, which relates the density limit with the quality of the confinement and the specific characteristics of the radiator that are accounted for in the proportionality factor. The question whether such a limit is to be expected in a reactor environment, where the current edge radiators, carbon and oxygen, will not be present is in order. Tungsten as the main wall element displays a cooling factor with only mild temperature dependence. However, it is to be expected that tungsten levels are kept very low and extrinsic edge radiators are used to control power exhaust in the edge and divertor regions. These would also exhibit higher cooling rates below a critical temperature and, in this sense, a density limit scaling of the Sudo type could be relevant for reactor conditions.

Taking the scalings \eqref{eq:density}, \eqref{eq:BRrelation} and \eqref{eq:nc} it easy to show that
\begin{equation}
n/n_c \sim P_h^{-0.57}B^{-0.01}\;.
\end{equation}
 So, for constant fusion power and gain and a density scaling as  $n\sim B^2$, the ratio $n/n_c$ remains constant as shown in figure \ref{fig:density}. This density scaling is therefore consistent with the presently known stellarator density limit.

\section*{Acknowledgements}
We are thankful to C. D. Beidler, H. Laqua, T. Sunn Pedersen, F. Warmer, R. Wolf and H. Zohm for reading the manuscript and providing useful comments. We thank F. Parra for discussions on the content of section \ref{sec:transport}, \rev{P. Helander for suggesting the study of increased D/T fraction of section \ref{sec:Bovereng} and J. de la Riva for his help implementing changes in the code}. The first author acknowledges the sustained interaction with C. D.  Beidler, A. Dinklage and F. Warmer on stellarator physics and reactor studies.

\printbibliography
\end{document}